\documentclass[preprints,article,accept,moreauthors,pdftex]{Definitions/mdpi} 
\usepackage{tipa}

\newcommand{\phat}[1]{\hat{#1}}
\firstpage{1} 
\pubvolume{xx}
\issuenum{1}
\articlenumber{5}
\pubyear{2019}
\copyrightyear{2019}
\history{Received: 6th August 2019; Accepted: 10th September 2019; Published: 16th September 2019}

\Title{Alternative Visual Units for an Optimized Phoneme-Based Lipreading System}

\Author{Helen L. Bear 
 $^{1,}$*\orcidA{} and Richard Harvey $^{2}$\orcidB{}}
\AuthorNames{Helen L. Bear and Richard Harvey}

\address{%
$^{1}$ \quad School of Electronic Engineering and Computer Science, Queen Mary University of London, London, E1 4NS, UK\\
$^{2}$ \quad School of Computing Sciences, University of East Anglia, Norwich, NR4 7TJ, UK; r.w.harvey@uea.ac.uk}

\corres{Correspondence: h.bear@qmul.ac.uk}

\conference{} 

\abstract{Lipreading is understanding speech from observed lip movements. An observed series of lip motions is an ordered sequence of visual lip gestures. These gestures are commonly known, but as yet are not formally defined, as `visemes'. In this article, we describe a structured approach which allows us to create speaker-dependent visemes with a fixed number of visemes within each set. We create sets of visemes for sizes two to 45. Each set of visemes is based upon clustering phonemes, thus each set has a unique phoneme-to-viseme mapping.
We first present an experiment using these maps and the Resource Management Audio-Visual (RMAV) dataset which shows the effect of changing the viseme map size in speaker-dependent machine lipreading and demonstrate that word recognition with phoneme classifiers is possible. Furthermore, we show that there are intermediate units between visemes and phonemes which are better still. Second, we present a novel two-pass training scheme for phoneme classifiers. This approach  uses our new intermediary visual units from our first experiment in the first pass as classifiers; before using the phoneme-to-viseme maps, we retrain these into phoneme classifiers. This method significantly improves on previous lipreading results with RMAV speakers. }


\dataset{RMAV Active Appearance Model Features can be found at http://doi.org/10.5281/zenodo.2576567}

\begin{document}


\section{Introduction}
The concept of phonemes is well developed in speech recognition and derives from a definition in phonetics as ``the smallest sound one can articulate'' \cite{international1999handbook}. Phonemes are analogous to atoms---they are the building blocks of speech. While they are an approximation, in~practice that approximation has been remarkably robust~\cite{hinton2012deep}. Not only are phonemes used by linguists and audiologists to describe speech, they are widely used in large-vocabulary speech recognition as the acoustic classes, or~`units', to~be recognized~\cite{hinton2012deep,wollmer2010recognition, triefenbach2010phoneme}. Sequences of unit estimates can be strung together to infer words and~sentences.

Comprehending visual speech, or~lipreading, is much less well developed~\cite{bear2016decoding}. The~units considered to be equivalent to phonemes are called visemes~\cite{cappelletta2011viseme} but, even in English, there is no clear agreement on the visemes~\cite{goldschen1996rationale}, and~in~\cite{bear2017phoneme} for example, it is noted that there are at least $120$ proposed viseme sets. This large number arises because some authors take vowels~\cite{montgomery1983physical}, and~others consonants~\cite{walden1977effects}, but~also because, of~the proposed sets, some are derived from linguistic principles~\cite{neti2000audio,woodward1960phoneme}, some are the results of human lipreading experiments~\cite{fisher1968confusions, finn1988automatic}, others are data-derived~\cite{bear2017phoneme, lee2002audio}, and~others still are hybrids of these approaches~\cite{bozkurt2007comparison}. 

Despite the challenges, a~number of lipreading systems have been built using visemes (\cite{shaikh2010lip,bear2014resolution} for example). When building a viseme recognizer a complication is that multiple phonemes will map onto a single viseme~\cite{bear2017phoneme}. A~common example is the $/p/$, $/b/$, and~$/m/$ bilabial sounds which are often grouped into one viseme~\cite{binnie1976visual, disney, lip_reading18}. Attempts to draw mappings between the phonemes and visemes have been tested~\cite{bear2017phoneme,bear2014some} but to date these mappings have not yet proven to improve machine lipreading~significantly.
 
On the other hand, there is an emerging body of work~\cite{thangthai2017comparing, howell2013confusion} that, despite the caveats above, is demonstrating that phoneme lipreading systems can outperform viseme recognizers. In~essence it is a tradeoff: does one use viseme units which are tuned to the shape of the lips but suffer with inaccuracies caused by visual confusions between words that sound different but look identical~\cite{thangthai2017comparing}; or does one stick to phonetic units knowing that many of the phonemes are difficult to distinguish on the lips?

These visual confusions are called homophenes~\cite{10.1080/00221309}. 
We demonstrate the homophenous word difficulty, with~some examples in Table~\ref{tab:homophones} from~\cite{thangthai2017comparing}. In~this example, Jeffers visemes~\cite{jeffers1971speechreading} have been used to translate the phonemes into viseme~strings.
 
\begin{table}[H]
\centering
\caption{Example of phoneme and viseme dictionary with its corresponding IPA symbols~\cite{thangthai2017comparing}.}
\begin{tabular}{lcc}
\toprule
\textbf{Word Entry} & \textbf{Phoneme Dictionary} & \textbf{Viseme Dictionary} \\ 
\midrule
TALK	&   /t/ /\textopeno/ /k/         		& /C/ /V1/ /H/            \\ 
TONGUE 	&   /t/ /\textturnv/ /\textipa{N}/ 	& /C/ /V1/ /H/           \\ 
DOG 	&   /d/ \textopeno/ /g/       		& /C/ /V1/ /H/            \\ 
DUG 	&   /d/ /\textturnv/ /g/        		& /C/ /V1/ /H/            \\ 
\cmidrule{1-3} 
CARE  	&   /k/ /e/ /r/				& /H/ /V3/ /A/            \\ 
WELL  	&   /w/ /e/ /l/  			      	& /H/ /V3/ /A/            \\
WHERE 	&   /w/ /e/ /r/         			& /H/ /V3/ /A/            \\ 
WEAR  	&   /w/ /e/ /r/         			& /H/ /V3/ /A/           \\
WHILE 	&   /w/ /ai/ /l/        			& /H/ /V3/ /A/            \\ 
\bottomrule
\end{tabular}
\label{tab:homophones}
\end{table}

However, as~we shall show in this paper, it need not be an either/or approach to phonemes or visemes; we develop a novel method that allows us to vary the number of classes/visual units. 
This~means we can tune the visual units as an intermediary state between the visual and audio spaces and we can also optimize against the competing trends of homopheneiosity~\cite{bear2017visual,bear2018comparing} and accuracy~\cite{htk34}. Thus, in this work, we use the term visemes for the traditional visemes, and~the term visual units for our new intermediary units which we propose will improve phoneme~classifiers. 

We are motivated in our work because lipreading is a difficult challenge from speech signals. Speech signals are bimodal (that is they have two channels of information, audio and visual) and significant prior work uses both. For~example~\cite{ngiam2011multimodal} uses audio-visual speech recognition to demonstrate cross modality learning. However in our case, lipreading, which is useful for understanding speech when audio speech is too noisy to recognize easily, is classifying speech from only the visual information channel in speech signals thus, as~we shall present, we use a novel training method which uses new visual units and phonemes in a complimentary~fashion. 

This paper is an extended version of our prior work~\cite{bear2015findingphonemes,bear2016decoding}, this work is relevant to all classifiers since the choice of visual unit matters and is made before the classifier is trained. In~other words, the~choice of visual units must be made early in the design process and a non-optimal choice can be very expensive in terms of~performance. 

The rest of this paper is structured as follows; we summarize prior viseme research for lipreading by both humans and machines, and~describe the state-of-the-art approaches for lipreading systems in a background section. Then we present an experiment in which we demonstrate how we can find the optimal number of visual units within a set; this is an essential preliminary test to define the scope of the second task. We present the data for all experiments within this section. The~preliminary test includes phoneme classification and clustering for new visual unit generation before analyzing the results to find the optimal visual unit~sets. 

These optimal visual unit sets are used to test our  novel method for training phoneme-labeled classifiers by using these sets as an initialization stage in the training phase of a conventional lipreading system. As~part of this second task, we also present a side task of deducing the right units for lipreading language models used in the lipreading system. Finally, we present the results of the new training method and draw conclusions before suggesting future work. Thus, we have three main contributions: 
\begin{itemize}[leftmargin=2.3em,labelsep=6mm]
    \item a method for finding optimal visual units, 
    \item a review of language model units for lipreading systems,
    \item a new training paradigm for lipreading systems.
\end{itemize}

\section{Background}
Table~\ref{tab:Confusion_Factors} summarizes the most common viseme sets in the literature used for both human and machine lip reading. 
The range of set sizes is from four (Woodward~\cite{woodward1960phoneme}) to 21 (Nichie~\cite{lip_reading18}). Note that not all viseme sets represent the same number of phonemes. Furthermore some of these use American English and others British English so there are minor variations in the phoneme sets. (American English phonemes tend to use diacritics~\cite{labov2005atlas}.)

\begin{table}[H] 
\centering 
\caption{Ratio of visemes to phonemes in previous viseme sets from literature.} 
\begin{tabular}{ l  r  l  r } 
\toprule
\textbf{Set} & \textbf{V:P} & \textbf{Set} & \textbf{V:P} \\ 
\midrule
Woodward~\cite{woodward1960phoneme} & 4:24 & Fisher~\cite{fisher1968confusions} & 5:21 \\ 
Lee~\cite{lee2002audio} & 9:38 & Jeffers~\cite{jeffers1971speechreading}  & 11:42  \\ 
Neti~\cite{neti2000audio} & 12:43 & Franks~\cite{franks1972confusion}& 5:17 \\ 
Disney~\cite{disney} & 10:33 & Kricos~\cite{kricos1982differences} & 8:24 \\
Hazen~\cite{Hazen1027972} & 14:39 & Bozkurt~\cite{bozkurt2007comparison} & 15:41 \\
Montgomery~\cite{montgomery1983physical} & 8:19 & Finn~\cite{finn1988automatic}& 10:23 \\ 
Nichie~\cite{lip_reading18} & 21:48 & Walden~\cite{walden1977effects} & 9:20  \\ 
\bottomrule 
\end{tabular} 
\label{tab:Confusion_Factors} 
\end{table}
Lipreading systems can be built with a range of architectures. Conventional systems are adopted from acoustic methods, often using Hidden Markov Models, for example as in~\cite{potamianos1998image}. More modern systems exploit deep learning methods~\cite{petridis2017end, stafylakis2017combining}. Deep learning has been deployed in two configurations: (i) as a replacement for the GMM in the Hidden Markov Models (HMM) and (ii) in a configuration known as end-to-end~learning. 

However, the~high-level architectures have similarities: first the face of the speaker must be tracked or located; then some form of features are extracted; then a classification model is trained and tested on unseen data, optionally using a language model to improve the classification output (e.g.~\cite{le2017generating}). Throughout this process one must translate between the words spoken (and captured in the training videos), to~their phonetic pronunciation, to~their visual representation on the lips, and~back again for a useful transcript. 



\section{Finding a Robust Range of Intermediate Visual~Units}
In our first example we use the RMAV dataset~\cite{improveVis} and the BEEP pronunciation dictionary~\cite{beep}. \textls[-15]{Figure~\ref{fig:process} shows a high-level overview of the first task. We begin with classification using phoneme-labeled} classifiers. The~output of this task is a set of speaker-dependent confusion matrices. The~data in these are used to cluster together single phonemes (monophones) into subgroups of visual units, based~upon~confusions. 
 
However, conversely to the approach in~\cite{bear2017phoneme} we implement an alternative phoneme clustering process (described in detail in Section~\ref{sec:newclusteringalg}). The~key difference between the ad-hoc viseme choices compared in~\cite{bear2017phoneme} and our new clustering approach, is our ability to choose the number of visual units, whereas in prior viseme sets, this is~fixed.

With our new algorithm, we create a new phoneme-to-viseme (P2V) mapping every time a pair of classes is re-classified into a new class, thus reducing the number of classes in a set by one each time. In~the phonetic transcripts of our $12$ speakers, there is a maximum of $45$ phonemes, therefore we can create at most $45$ P2V maps for each speaker. We note that the real number of maps we can derive depends upon the number of phonemes classified during step one of Figure~\ref{fig:process}. During~this preliminary phoneme classification, should a phoneme not be classified, either incorrectly or correctly, then it is an omission in the confusion matrix from which our visual units are created. Thus, we have \emph{up to} $45$ sets of visual unit labels per speaker with which to label our~classifiers. 

\begin{figure}[H] 
\centering 
\includegraphics[width=0.5\textwidth]{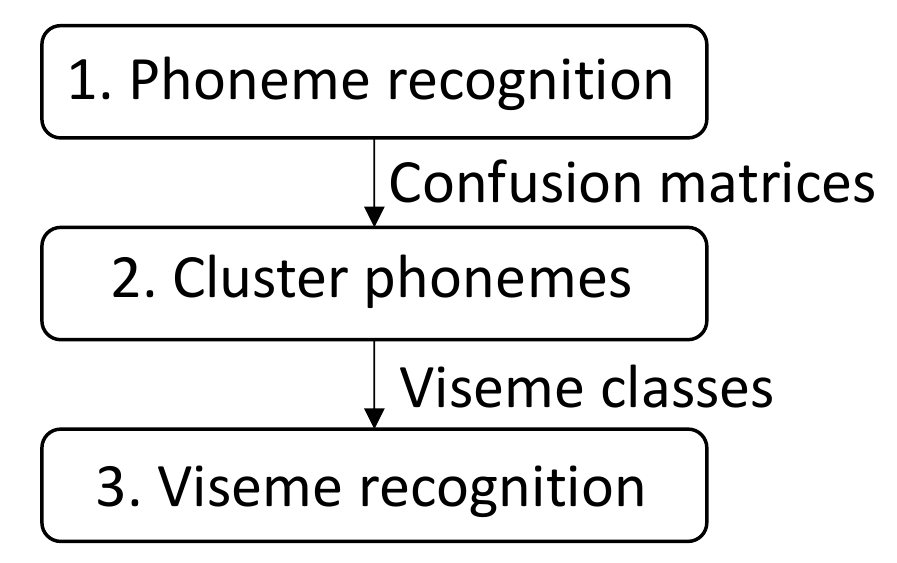}
\caption{Three-step high-level process for visual unit classification where the visual units are derived from phoneme~confusions.} 
\label{fig:process} 
\end{figure} 
There is the option to measure performance using phoneme, viseme, or~word error. Here we choose word error~\cite{bearTaylor} because viseme error varies as the number of visemes varies which leads to unfair comparisons and phoneme error is not as close to what we believe to be of interest to users which is transcript~error.  
 
\subsection{Data}
The RMAV dataset (formerly known as LiLIR) 
consists of $20$ British English speakers (we use the 12 speakers who had tracked features available; seven male and five female) and up to $200$ utterances per speaker of the Resource Management (RM) sentences which totals between $1362$ and $1802$ words each. The~sentences selected for the RMAV speakers are a subset of the full RM dataset~\cite{fisher1986darpa} transcripts. They were selected to maintain as much coverage of all phonemes as possible as shown in Figure~\ref{fig:histogram} and realistic to English conversation~\cite{improveVis}. The~original videos were recorded in high definition ($1920 \times 1080$) and in a full-frontal position at 25 fs$^{-1}$. Individual speakers are tracked using Linear Predictors~\cite{ong2011robust} and Active Appearance Model~\cite{Matthews_Baker_2004} features of concatenated shape and appearance information have been~extracted. 

\begin{figure}[H] 
\centering
\includegraphics[width=.75\textwidth]{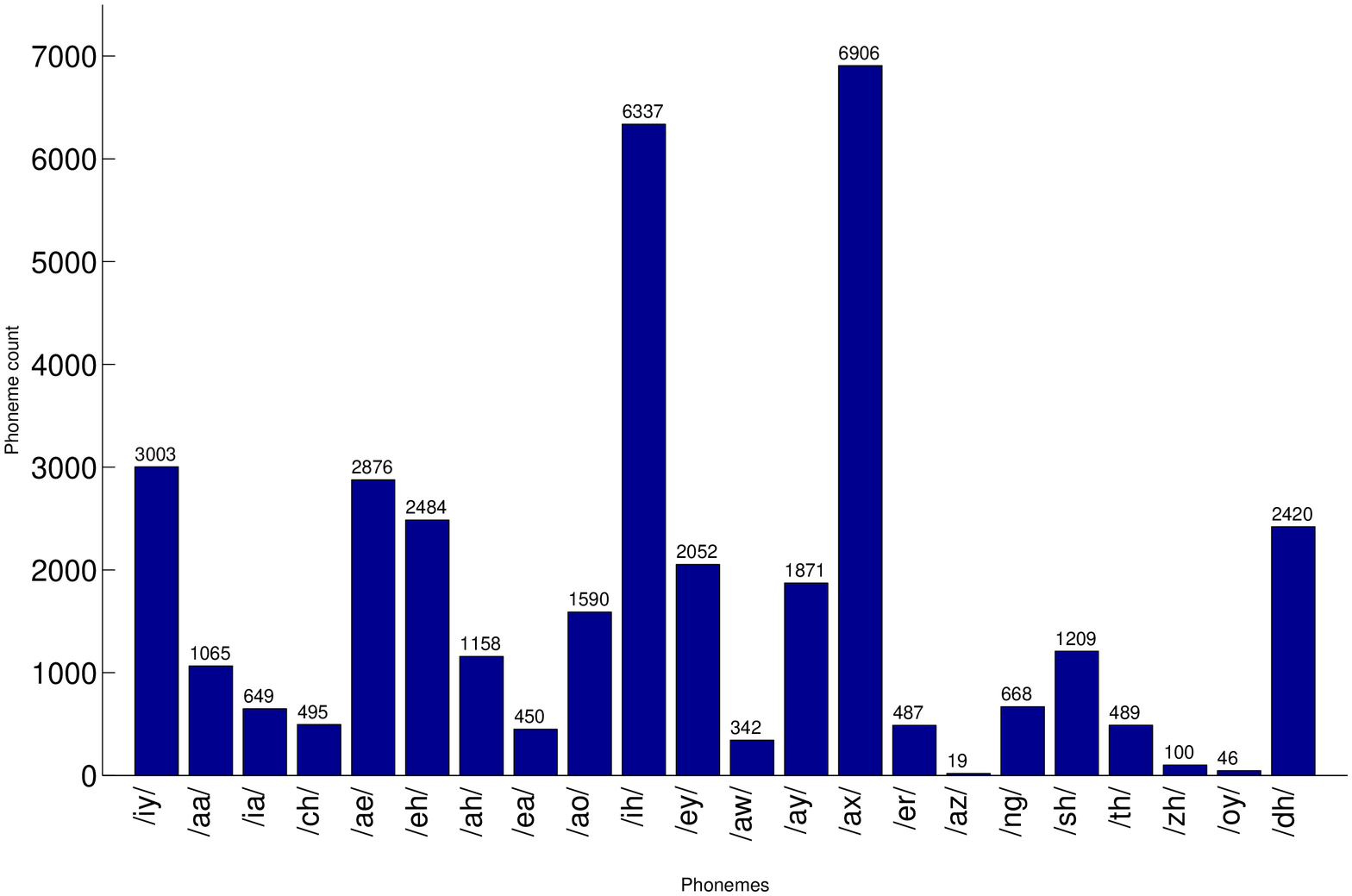}
\caption{\textit{Cont.}}
\label{fig:histogram}
\end{figure}

\begin{figure}[H]\ContinuedFloat
\centering
\includegraphics[width=.75\textwidth]{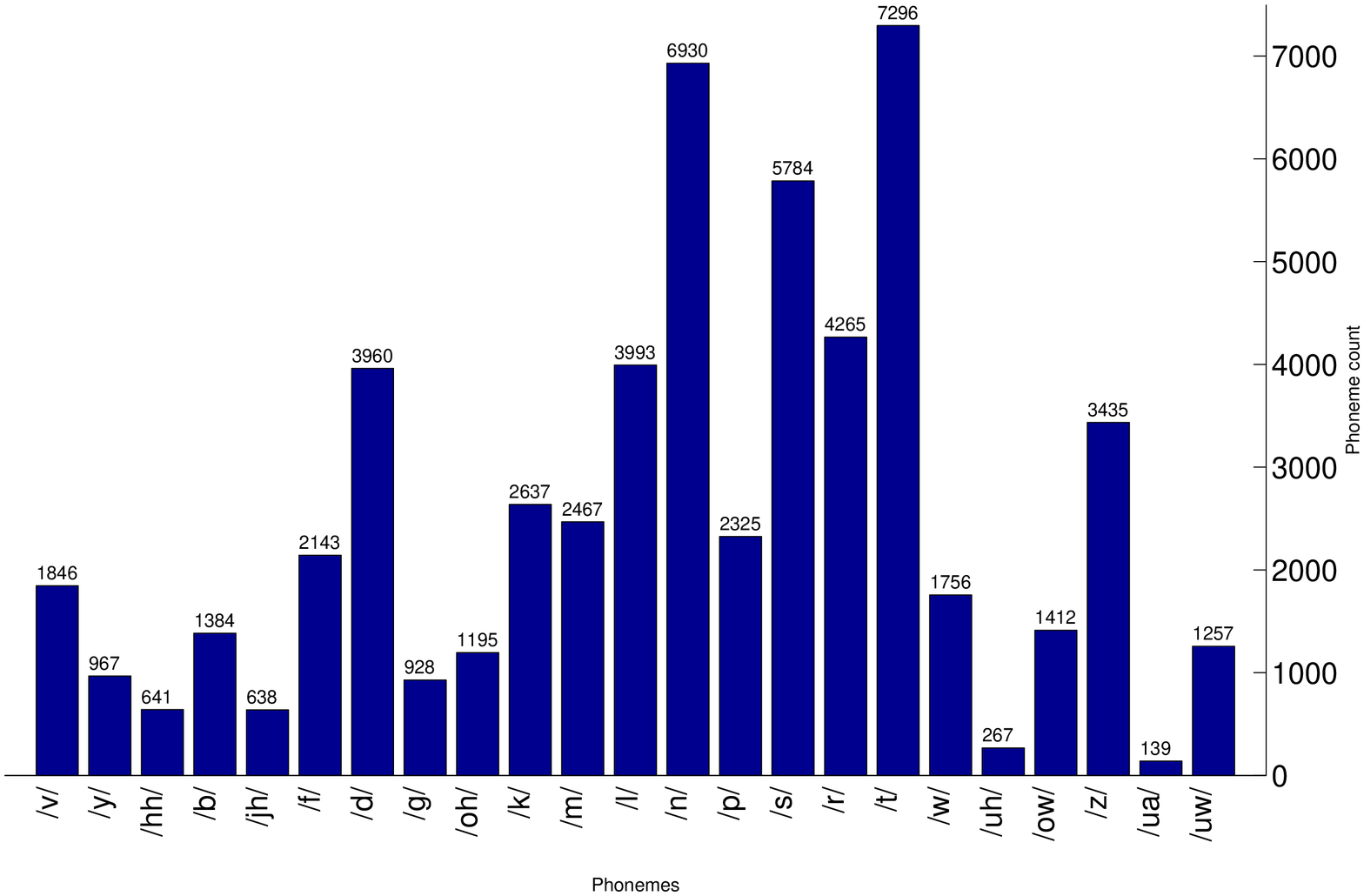}
\caption{Occurrence frequency of phonemes in the RMAV~dataset.} 
\label{fig:histogram} 
\end{figure}
\unskip 

\subsection{Linear Predictor~Tracking}
Linear Predictors (LP) are a person-specific and data-driven facial tracking method. Devised primarily for observing visual changes in the face during speech, these make it possible to cope with facial feature configurations not present in the training data by treating each feature~independently. 
 
The linear predictor is the central point around which support pixels are used to identify the change in position of the central point over time. The~central point is observed as a landmark on the outline of a feature. In~this method both the shape (comprised of landmarks) and the pixel information surrounding the linear predictor position are intrinsically linked. Linear predictors have been successfully used to track objects in motion, for~example~\cite{matas2006learning}. 

\subsection{Active Appearance Model~Features}
 
\begin{figure}[H] 
\centering 
\includegraphics[width=.33\textwidth]{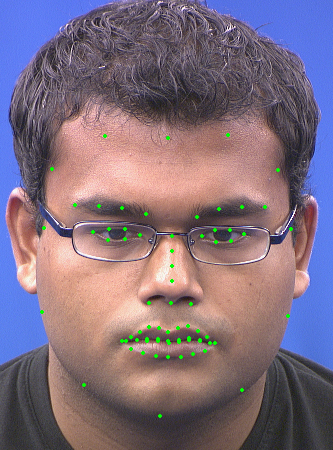} 
\includegraphics[width=.33\textwidth]{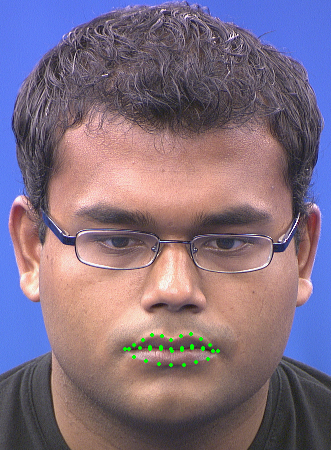}
\caption{Landmarks in a full-face AAM used to track a face (\textbf{left}) and the lip-only AAM landmarks (\textbf{right})for feature extraction.}  
\label{fig:landmarks} 
\end{figure}

AAM features~\cite{Matthews_Baker_2004} of concatenated shape and appearance information have been extracted. We~track using a full-face model (Figure~\ref{fig:landmarks} (left)) but the final features are reduced to information from the lip area alone (Figure~\ref{fig:landmarks} (right)). Shape features (\ref{eq:shapecombined}) are based solely upon the lip shape and positioning during the duration of the speaker speaking. The~landmark positions can be compactly represented using a linear model of the form:
\begin{equation}
s = s_0 + \sum_{i=1}^ms_ip_i
\label{eq:shapecombined}
\end{equation}
where $s_0$ is the mean shape and $s_i$ are the modes. The~appearance features are computed over pixels, the~original images having been warped to the mean shape. So $A_0(x)$ is the mean appearance and appearance is described as a sum over modal appearances:
\begin{equation} 
A(x) = A_0(x) + \sum_{i=1}^l{\lambda}_iA_i(x) \qquad \forall x \in S_0
\label{eq:appcombined}
\end{equation} 

Combined features are the concatenation of shape and appearance after PCA has been applied to each independently. The~AAM parameters for each speaker is in Table~\ref{tab:parameterslilirfeatures} (MATLAB files containing the extracted features can be downloaded from \url{http://zenodo.org/record/2576567}).

\unskip

 \begin{table}[H] 
\centering 
\caption{The number of parameters of shape, appearance, and~combined shape and appearance AAM features for the RMAV dataset speakers. Features retain 95\% variance of facial~information.}

\begin{tabular}{ l  r  r  r } 
\toprule
\textbf{Speaker}	& \textbf{Shape} & \textbf{Appearance} & \textbf{Combined} \\ 
\midrule 
S1	& 13 & 46	& 59 \\ 
S2 	& 13 & 47 & 60 \\ 
S3	& 13 & 43	& 56 \\ 
S4	& 13 & 47	& 60 \\ 
S5 	& 13 & 45 & 58 \\ 
S6	& 13 & 47	& 60 \\ 
S7 	& 13 & 37 & 50 \\ 
S8 	& 13 & 46 & 59 \\ 
S9	& 13 & 45	& 58 \\ 
S10 	& 13 & 45 & 58 \\ 
S11 	& 14 & 72 & 86 \\ 
S12	& 13 & 45	& 58 \\ 
\bottomrule
\end{tabular} 
\label{tab:parameterslilirfeatures}
 \end{table}
\unskip

\section{Clustering} 
\label{sec:newclusteringalg}
\unskip
\subsection{Step One: Phoneme~Classification} 
\label{sec:one} 
 To complete our preliminary phoneme classification, we implement $10$-fold cross-validation with replacement~\cite{efron1983leisurely}, over~the 200 sentences per speaker. This means $20$ test samples are randomly selected and omitted from training sample folds. Our classifiers are based upon Hidden Markov Models (HMMs) \cite{holmes2001speech} and implemented with the HTK toolkit~\cite{htk34}. We use the HTK tools as follows;
 \begin{enumerate}[leftmargin=*,labelsep=4.9mm]
 \item \texttt{HLed} creates our phoneme transcripts to be used as ground truth transcriptions. 
 \item \texttt{HCompV} initializes the HMMs using a `flat-start'~\cite{forney1973viterbi} using manually made prototype files for each speaker based upon their AAM parameters (listed in Table~\ref{tab:parameterslilirfeatures}) and the desired HMM parameters. The~prototype HMM is based upon a Gaussian mixture of five components and three state HMMs as per the work of~\cite{982900}.
 \item Using \texttt{HERest} we train the classifiers by re-estimating the HMM parameters $11$ times over via embedded training with the Baum-Welch algorithm~\cite{welch2003hidden}, more than 11 iterations and the HMM's overfit. Our list of HMMs includes a single-state, short-pause model, labeled $/sp/$ to model the short silences between words in the spoken sentences. States are tied with \texttt{HHed}. 
 \item We build a bigram word lattice using \texttt{HLStats} and \texttt{HBuild} and use this lattice to complete recognition with \texttt{HVite}. \texttt{HVite} uses our trained set of phoneme-labeled HMM classifiers to estimate what our test samples should be.
 \item The output transcripts from \texttt{HVite} are used with our ground truth transcriptions from \texttt{HLEd} as inputs into \texttt{HResults} to produce confusion matrices and lipreading accuracy scores. \texttt{HResults} uses an optimal string match using dynamic programming~\cite{htk34} to compare the ground truths with the prediction transcripts. 
 \end{enumerate}

 \subsection{Step Two: Phoneme~Clustering} 
\label{sec:two} 
Now we have our phoneme confusions (an example matrix is in Figure~\ref{table:examplecm}), we have ten confusion matrices per speaker (one for each fold of the cross-validation). We cluster the $m$ phonemes into new visual unit classes, one iteration at a~time. 

\begin{figure}[H]
\centering
\resizebox{\columnwidth}{!}{%
\begin{tabular}{|l l|c|c|c|c|c|c|c|c|c|c|c|c|c|c|c|}
\toprule
& & \multicolumn{15}{c|}{Predicted classes} \\
& \multicolumn{1}{r|}{} & /ae/ & /ay/ & /b/ & /c/ & /d/ & /ea/ & /f/ & /iy/ & /l/ & /m/ & /n/ & /oy/ & /p/ & /s/ & /t/  \\
\midrule
\multirow{16}{*}{Actual classes} & /ae/ & 76 & 2  & 1  & 5   & 2  & 1  & 3  & 1  & 1  & 3   & 1  & 5  & 0  & 0  & 4 \\ \cmidrule{2-17}
& /ay/ &  0  & 28 & 0  & 1   & 0  & 2  & 0  & 0  & 1  & 0   & 0  & 0  & 0  & 0  & 2  \\ \cmidrule{2-17}
& /b/ &  0  & 4  & 17 & 0   & 1  & 2  & 0  & 0  & 1  & 0   & 0  & 1  & 0  & 0  & 0  \\ \cmidrule{2-17}
& /c/ &  3  & 6  & 6  & 163 & 3  & 7  & 7  & 2  & 8  & 7   & 1  & 4  & 2  & 0  & 1  \\ \cmidrule{2-17}
& /d/ &  4  & 2  & 2  & 3   & 33 & 0  & 0  & 1  & 3  & 0   & 1  & 2  & 1  & 0  & 1  \\ \cmidrule{2-17}
& /ea/ &  2  & 0  & 0  & 6   & 1  & 9  & 0  & 0  & 1  & 0   & 0  & 0  & 0  & 0  & 0  \\ \cmidrule{2-17}
& /f/ &  4  & 1  & 0  & 3   & 1  & 1  & 40 & 0  & 0  & 1   & 5  & 2  & 0  & 0  & 0  \\ \cmidrule{2-17}
& /iy/ &  0  & 3  & 2  & 1   & 2  & 0  & 0  & 11 & 8  & 2   & 0  & 2  & 0  & 0  & 1  \\ \cmidrule{2-17}
& /l/ &  0  & 0  & 1  & 4   & 1  & 0  & 1  & 2  & 97 & 3   & 1  & 0  & 0  & 0  & 0  \\ \cmidrule{2-17}
& /m/ &  2  & 1  & 4  & 1   & 2  & 3  & 0  & 1  & 6  & 110 & 8  & 0  & 2  & 0  & 0  \\ \cmidrule{2-17}
& /n/ &  0  & 1  & 0  & 1   & 0  & 1  & 2  & 0  & 0  & 1   & 14 & 1  & 2  & 0  & 0  \\ \cmidrule{2-17}
& /oy/ &  0  & 0  & 0  & 3   & 1  & 1  & 4  & 1  & 1  & 3   & 1  & 16 & 1  & 0  & 0  \\ \cmidrule{2-17}
& /p/ &  0  & 0  & 0  & 0   & 0  & 0  & 0  & 0  & 0  & 0   & 1  & 0  & 3  & 0  & 0  \\ \cmidrule{2-17}
& /s/ &  0  & 0  & 0  & 0   & 0  & 0  & 0  & 0  & 0  & 0   & 0  & 0  & 0  & 84 & 0  \\ \cmidrule{2-17}
& /t/ & 1  & 3  & 0  & 2   & 1  & 1  & 1  & 0  & 0  & 1   & 2  & 0  & 0  & 0  & 28 \\  
\bottomrule
\end{tabular}%
}
\caption{An example phoneme confusion matrix.}
\label{table:examplecm}
\end{figure}

First we sum all ten matrices into one matrix to represent all the confusions for each speaker. Our clustering begins with this single specific speaker confusion matrix.
\begin{equation} 
[K_{m}]_{ij} = N (\phat{p}_j | p_i)\quad 
\label{eq3} 
\end{equation} 
where the $ij^{th}$ element is the count of the number of times phoneme $i$ is classified as phoneme $j$. This algorithm works with the column normalized version,
\begin{equation} 
[P_m]_{ij} = Pr\{p_i | \phat{p}_j \} \quad 
\label{eq4} 
\end{equation} 
the probability that given a classification of $p_j$ that the phoneme really was $p_i$. Merging of phonemes is done by looking for the two most confused phonemes and hence creating new matrices $K_{m-1}, P_{m-1}$. 
Specifically, for each possible merged pair a score, $q$, is calculated as:
\begin{equation} 
q = [P_{m}]_{rs} + [P_{m}]_{sr} \quad \\ 
= Pr\{\phat{P}r | Ps \} + Pr\{\phat{P}s | Pr\} 
\label{eq5} 
\end{equation} 
Vowels and consonants cannot be mixed, the~significant negative effect of mixing vowel and consonant phonemes in visemes was demonstrated in~\cite{bear2017phoneme}, so phonemes are assigned to one of two classes, $V$ or $C$, for~vowels and consonants respectively. The~pair with the highest $q$ is merged. We break equal scores randomly. This process is repeated until $m=2$. We stop at two because at this point we have two single classes, one class containing vowel phonemes, and~a second class of consonant phonemes. Each~intermediate step, $M = 45,44,43 ... 2$ forms another set of prospective visual units. 
 An example P2V mapping is shown in Table~\ref{tab:example} for RMAV speaker number one with ten visual~units. 
\begin{table}[H]
\centering 
\caption{An example P2V map, (for RMAV Speaker 1 with ten visual units).} 
\begin{tabular}{ll} 
\toprule 
\textbf{Visual Unit} & \textbf{Phonemes} \\ 
\midrule
$/v01/$ & /ax/ \\ 
$/v02/$ & /v/ \\ 
$/v03/$ & /\textopeno\textsci/ \\ 
$/v04/$ & /f/ /\textipa{Z}/ /w/ \\ 
$/v05/$ & /k/ /b/ /d/ /\textipa{T}/ /p/ \\ 
$/v06/$ & /l/ /d\textipa{Z}/ \\ 
$/v07/$ & /g/ /m/ /z/ /y/ /t\textipa{S}/ /\textipa{D}/ /s/ /r/ /t/ /\textipa{S}/ \\ 
$/v08/$ & /n/ /hh/ /\textipa{N}/ \\ 
$/v09/$ & /\textipa{E}/ /ae/ /\textopeno/ /uw/ /\textturnscripta/ /\textsci\textschwa/ /ey/ /ua/ /\textrevepsilon/ \\ 
$/v10/$ & /ay/ /\textscripta/ /\textturnv/ /\textscripta\textupsilon/ /\textupsilon/ /\textschwa\textupsilon/ /\textsci/ /iy/ /\textschwa/ /eh/ \\ 
\bottomrule 
\end{tabular} 
\label{tab:example} 
\end{table}
\unskip 
 
\subsection{Step Three: Visual Unit~Classification} 
\label{sec:visRecog} 
Step three is similar to step one. We again complete $10$-fold cross-validation with replacement~\cite{efron1983leisurely} over the $200$ sentences for each speaker using the same folds as the prior steps to prevent mixing the training and test data. Again, $20$ test samples are randomly selected to be omitted from the training folds. Again, with the HTK toolkit, we build new sets of HMM classifiers. This time however, our~classifiers are labeled with the visual units we have just created in step~two. 

We have a python script which translates the phoneme transcripts from using \texttt{HLed} in step one and the P2V maps from step two, into~visual unit transcripts, one for each P2V map. For~each set of visual units, visual unit HMMs are flat-started (\texttt{HCompV}) with the same speaker specific HMM prototypes as before (Gaussian mixtures are uniform across prototypes), re-estimated $11$ times over with \texttt{HERest}. A~bigram word lattice supports classification including a grammar scale factor of $1.0$ (shown to be optimum in~\cite{howell2013confusion}) and a transition penalty of $0.5$. 
 
The important difference this time is that the visual unit classes are now used as classifier labels. By~using these sets of classes which have been shown in step one to be visually confusing on the lips, we now perform classification for each class set. In~total this is at most $44$ sets, where the smallest set is of two classes (one with all the vowel phonemes and the other all the consonant phonemes), and~the largest set is of $45$ classes with one phoneme in each---thus the largest set for each speaker is a repeat of the phoneme classification task but using only phonemes which were originally recognized (either correctly or incorrectly) in step~one. 
 
\section{Optimal Visual Unit Set~Sizes} 
\label{sec:search}
\begin{figure}[H] 
\centering 
\includegraphics[width=0.95\linewidth]{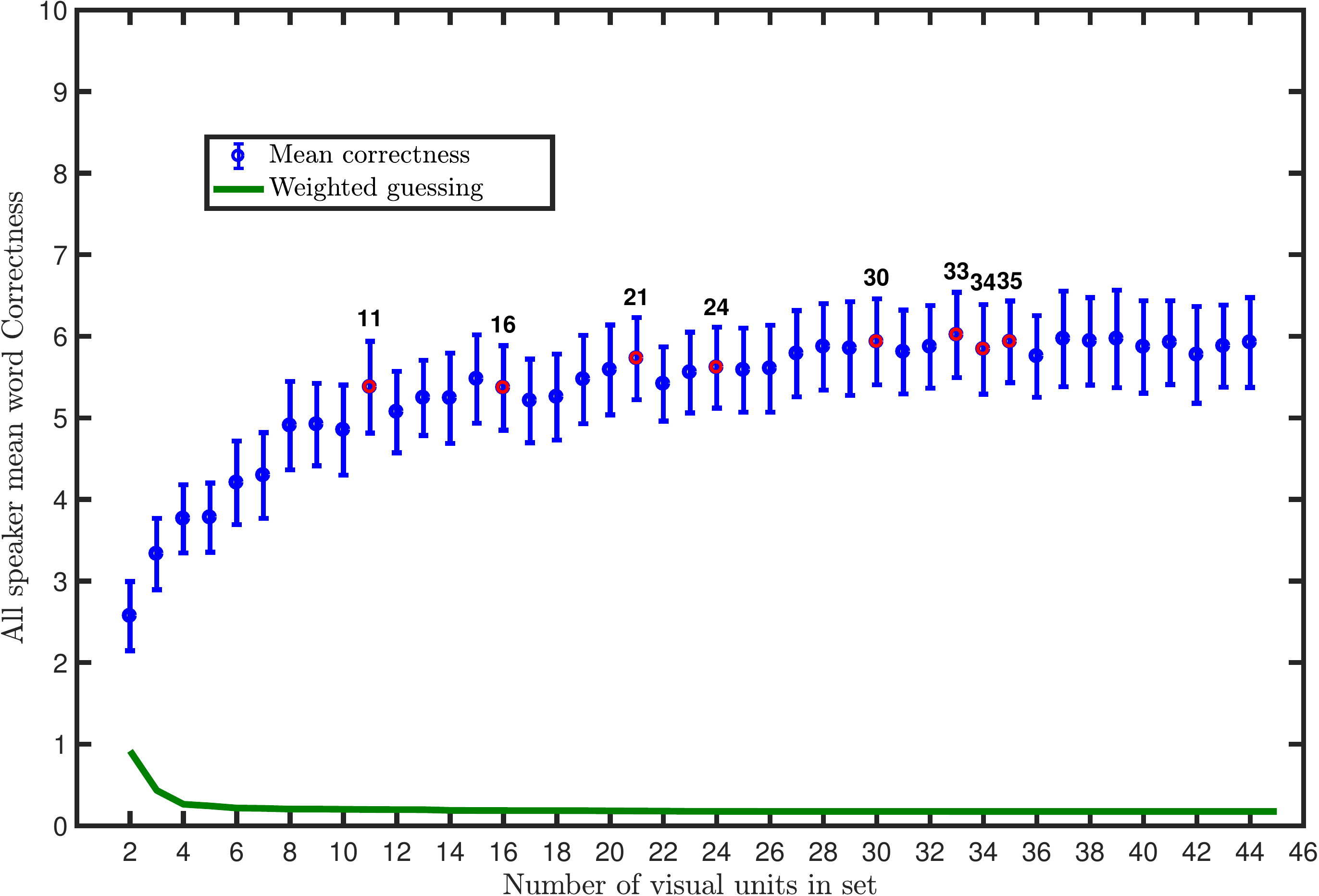} 
\caption{All-speaker mean word classification correctness $C\pm1se$.} 
\label{fig:correctness} 
\end{figure}

Figure~\ref{fig:correctness} plots word correctness on the $y$-axis for all $12$ speakers with error bars showing $\pm$ one standard error (se). The~$x$-axis shows the number of visual units. In~green we plot mean weighted guessing over all speakers for each viseme set. Individual speaker variations are in Appendix~\ref{sec:app}, Figures~\ref{fig:sp01}--\ref{fig:sp11}.

It is important in this case to weight the chance of guessing by visual homophenes as these vary by the size of the visual unit set. Visual unit sets which contain fewer visual units produce sequences of visual units which represent more than one word. These are homophenes. The~effect of homophenes can be seen on the left side of Figure~\ref{fig:correctness} and the graphs in Appendix~\ref{sec:app} with visual unit sets with fewer than $11$ visual units where homophenes become noticeable and language model can no longer correct these~confusions.
  
An example of a homophene in the RMAV data are the words `tonnes' and `since'. If~one uses Speaker 1's 10-visual unit P2V map, both words transcribe into visual units as `$/v7/$ $/v10/$ $/v8/$ $/v7/$'. In~practice a language model, or~word lattice, will tend to reduce such confusions since the lattice models the probability of word $N$-grams which means that probable combinations such as ``metric tonnes'' will be favored over ``metric since'' \cite{thangthai2017comparing}.
  

We see all our word correctness scores are significantly above guessing albeit still low. There~is variation between speakers, but~there is a clear overall trend. Superior performance is to be found with larger numbers of visual units. An~important point is some authors report viseme accuracy instead of word correctness~\cite{bearTaylor}. This is unhelpful as it masks the effect of homophenous words on  performance. Had we reported this then the positive effect of larger visual unit sets would not be~visible.

In Figure~\ref{fig:correctness} we highlight in red the class sets which, for~any speaker, have shown a significant classification improvement (with non-overlapping error bars) over the adjacent set of units on its right side along the $x$-axis. Error bars overlap once the correctness is averaged so Table~\ref{tab:merges} lists these combinations for each speaker. These red points show where we can identify the pairs of classes which, when merged into one class, significantly improve classification. If~we refer to the speaker demographic factors such as gender or age, we find no apparent pattern through these visual unit combinations. So, we have further evidence to reinforce the idea that all speakers have a unique visual speech signal,~\cite{607030}. In~\cite{bear2015speaker} this is suggested to be due to how the trajectory between visual units varies by speaker, due to such things as rate of speech~\cite{taylor2014effect}. This is how difficult finding a set of cross-speaker visual units can be when phonemes need alternative groupings for each individual~\cite{bear2017visual}.

\begin{table}[H] 
\centering 
\tablesize{\footnotesize} 

\caption{visual unit class merges which improve word classification in correctness; $V_n=V_i+V_j$.} 
   \resizebox{\columnwidth}{!}{%
\begin{tabular}{ l  l  l  l  l  l } 
\toprule
\textbf{Speaker} & \textbf{Set No} & \boldmath{$V_i$} & \boldmath{$V_j$} & \textbf{Set No} & \boldmath{$V_n$} \\ 
\midrule 
Sp01 & 35 & /s/ /r/ & /\textipa{D}/ & 34 & /s/ /r/ /\textipa{D}/ \\ 
Sp02 & 22 & /d/ & /z/ /y/ & 21 & /d/ /z/ /y/ \\ 
Sp03 & 34 & /b/ /t\textipa{S}/ & /\textipa{Z}/ & 33 & /b/ /t\textipa{S}/ /\textipa{Z}/ \\ 
Sp03 & 31 & /\textipa{Z}/ /b/ /t\textipa{S}/ & /z/ & 30 & /\textipa{Z}/ /b/ /t\textipa{S}/ /z/ \\ 
Sp03 & 25 & /p/ /r/ & /\textipa{N}/ & 24 & /p/ /r/ /\textipa{N}/ \\ 
Sp05 & 17 & /ae/ & /eh/ & 16 & /ae/ /eh/ \\ 
Sp06 & 35 & /ae/ /\textturnv/ & /iy/ & 34 & /ae/ /\textturnv/ /iy/ \\ 
Sp09 & 12 & /b/ /w/ /v/ & /d\textipa{Z}/ /hh/ & 11 & /b/ /w/ /v/ /d\textipa{Z}/ /hh/ \\ 
Sp12 & 36 & /\textturnv/ & /\textopeno/ & 35 & /\textturnv/ /\textopeno/ \\ 
\bottomrule 
\end{tabular} }%
\label{tab:merges} 
\end{table}
\unskip

\section{Discussion}
\label{sec:discussion}


In Figure~\ref{fig:correctness} we have plotted mean word correctness, $C$, over~all  $12$ speakers and weighted guessing ($1/(number Of Units$) in green. Here we see that within~one standard error, there is a monotonic trend. Small numbers of units perform worse than phonemes and which supports the claim that phonemes are preferred to visemes but, it would be an oversimplification to assert that higher accuracy lipreading can be achieved with phonemes as this has not been shown in our results with significance. Rather we say that, generally, visual unit sets with higher numbers of visual unit classes outperform the smaller sets. In~\cite{bear2017phoneme} the authors reviewed $120$ of previous phoneme-to-viseme (P2V) maps, typically these consist of between $10$ and $35$ visual units~\cite{bear2014phoneme}. For~example the Lee set consists of six consonant visemes and five vowel visemes~\cite{lee2002audio} and Jeffers~\cite{jeffers1971speechreading} group phonemes into eight vowel and three consonant~visemes. 

In Figures~\ref{fig:sp01}--\ref{fig:sp11} and Figure \ref{fig:correctness} we present a definite rapid decrease in lipreading word correctness for visemes sets containing fewer than ten visemes. However, positively, the~region visemes sets of sizes between $11$ and $20$ contain the optimum viseme set for three out of the 12 speakers which is more than random chance. This means, for~each speaker, we have found and presented an optimal number of visual units (shown by the best performing results in Figures~\ref{fig:sp01}--\ref{fig:sp11}) but the optimal number is not related to any of the conventional viseme definitions, nor is it consistent across speakers. Table~\ref{tab:pr_vals} shows the word correctness, $C_w$, of~each speakers phoneme~classification. 
\begin{table}[H] 
\centering 
\caption{Phoneme correctness $C$ for each speaker (right-hand data points of Figures~\ref{fig:sp01}--\ref{fig:sp11}).} 
\begin{tabular}{lrrrrrrrrrrrr} 
\toprule
Speaker & 1 & 2 & 3 & 4 & 5 & 6  \\ 
Phoneme $C$ & 0.05 & 0.06 & 0.06 & 0.05 & 0.06 & 0.06 \\ 
\midrule 
Speaker & 7 & 8 & 9 &10 & 11 & 12 \\ 
Phoneme $C$ & 0.06 & 0.06 & 0.06 & 0.07 & 0.06 & 0.06 \\ 
\bottomrule 
\end{tabular} 
\label{tab:pr_vals} 
\end{table}
\unskip

\section{Hierarchical Training for Weak-Learned Visual~Units} 
\label{chap:support network} 
Figure~\ref{fig:correctness} showed our first results derived using an adapted version of the algorithm described in~\cite{bear2014phoneme}. 
Table~\ref{tab:merges} also shows us, for~each of our $12$ speakers the significantly improving visual unit sets. These sets are those where one single change of visual unit grouping has resulted in a significant (greater than one standard error over ten folds) increase in word correctness. This tells us that there are some units between the traditional visemes (for example~\cite{fisher1968confusions, lip_reading18, disney}), and~phonemes which are better for visual speech~classification. 

Table~\ref{tab:merges} (\cite{bear2015findingphonemes}) shows us several significantly improving sets. Our suggestions for why these are interesting are; first the tradeoff of homophenes against accuracy. It is possible these are the groupings where the accuracy improvement is significantly improving, despite the extra homophenes created as the number of visual units in the set decreases. Either the increase in homophenes is negligible or, the~number of training samples for two visually indistinguishable classes significantly increases when~combined.


We propose a novel idea; to implement hierarchical classifier training using both visual units and phonemes in sequence. Some work in acoustic speech recognition has used this layered approach to model building with success e.g.~\cite{morgan2012deep}. It is our intention use our new range of visemes to test if our new training algorithm can improve phoneme classification without the need for more training data as this approach shares training data across models. This premise avoids the negative effects of introducing more homophenes because of the second layer of training discriminates between the sub-units\textls[-15]{ within the first layer. This will assist the identification of the more subtle but important differences in visual gestures representing alternative phonemes. We note from~\cite{bear2014some} that using the wrong clusters of phonemes is worse than using none, and~also that this new approach aims to optimize performance within the scope of the datasets and system affects described previously in Sections~\ref{sec:search} and~\ref{sec:discussion}. }

A bonus of our revised classification scheme is that because we weakly train the classifier before phoneme training, we remove any desire to consider post-processing methods (e.g. weighted finite state transducers~\cite{howell2013confusion}) to reverse the P2V mapping in order to decode the real phoneme~recognized.


In Figure~\ref{fig:correctness}, the~performance of classifiers with small numbers of visual units (fewer than $10$) is poor. As~described previously, we attributed this to the large number of homophenes. At~the other side of our figure, sets containing large numbers of visual units (greater than $35$) do not significantly, or~even noticeably, improve the correctness. This is where many phonetic variations are visually indistinguishable on the lips. Also taking into account the set numbers printed in black (which are the significantly improving visual unit sets) we focus on sets of visual units in the size range $11$ to $35$ with the same $12$ RMAV speakers for our experiments using hierarchical training of phoneme~classifiers.


\begin{figure}[H] 
\centering 
\includegraphics[width=0.9\textwidth]{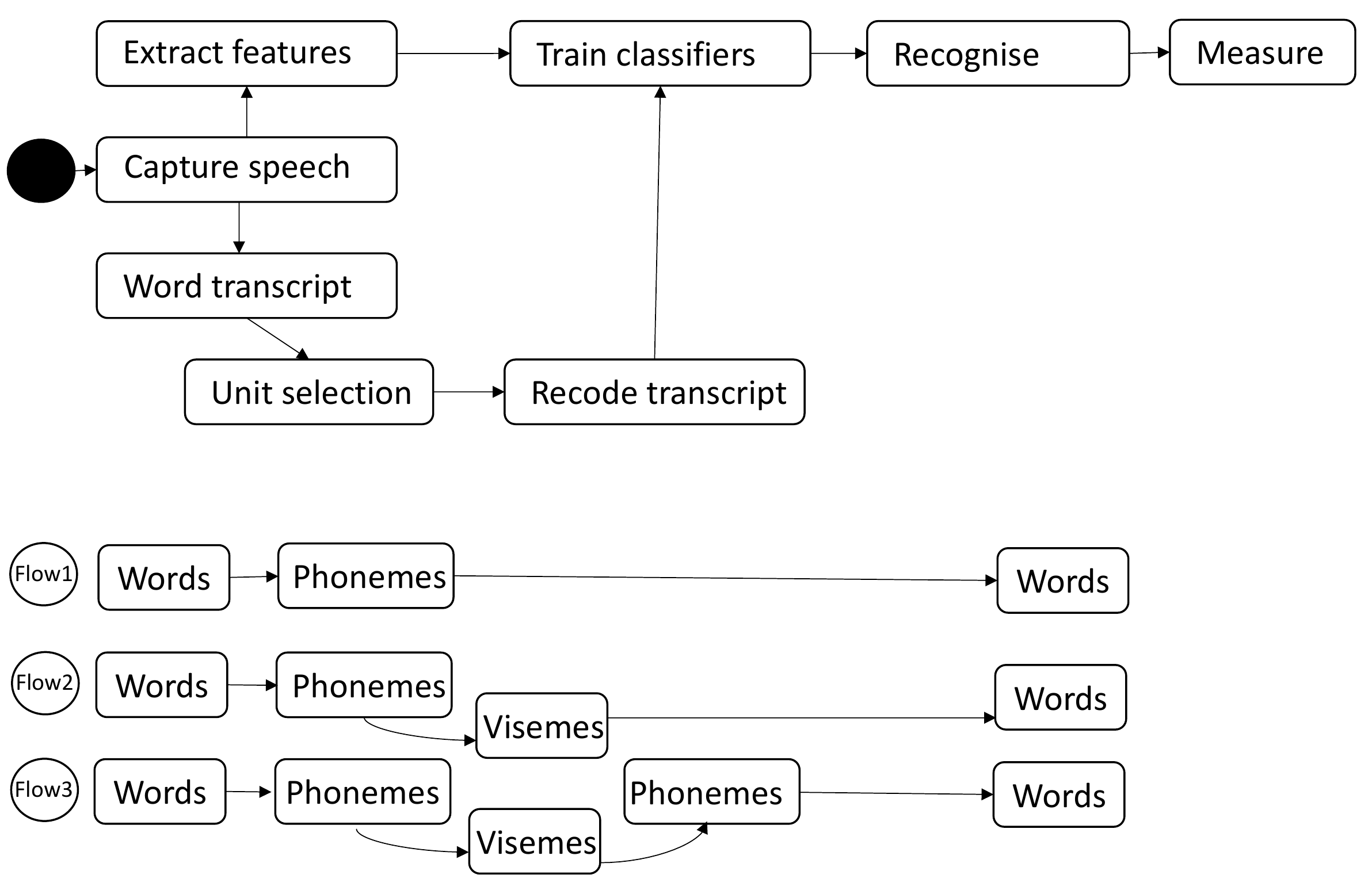}
\caption{\textls[-25]{\textbf{Top}: a high-level lipreading system, and~\textbf{Bottom}: where conversions between words, phonemes,} and~visual units can occur in lipreading systems in three different~flows.} 
\label{fig:unitphases}
\end{figure}

Here, we use our knowledge of visual speech to drive our novel redesign of the conventional training method. In Figure~\ref{fig:unitphases} shows how we make it earlier in the process. The~top of Figure~\ref{fig:unitphases} in black boxes shows the steps of a lipreading system, divided into phases where the units change from words, to~phonemes, to~visual units (where used). Flow 1 shows how we translate the word ground truth into phonemes using a pronunciation dictionary (e.g.~\cite{beep} or~\cite{cmudict}) for labeling the classifiers, before~decoding with a word language model. Flow 2 below this, using visual units. The~variation in flow 2 shows we translate from visual unit trained classifiers back into words using the word network. Finally, row three shows our new approach, where we introduce an extra step into the training phase, which means classifiers are initialized as visual units, before~retraining them into phoneme classifiers before word decoding. We describe this new process in detail~now. 

\section{Classifier Adaptation~Training}
\begin{figure}[H]
  \centering
\includegraphics[width=0.9\textwidth]{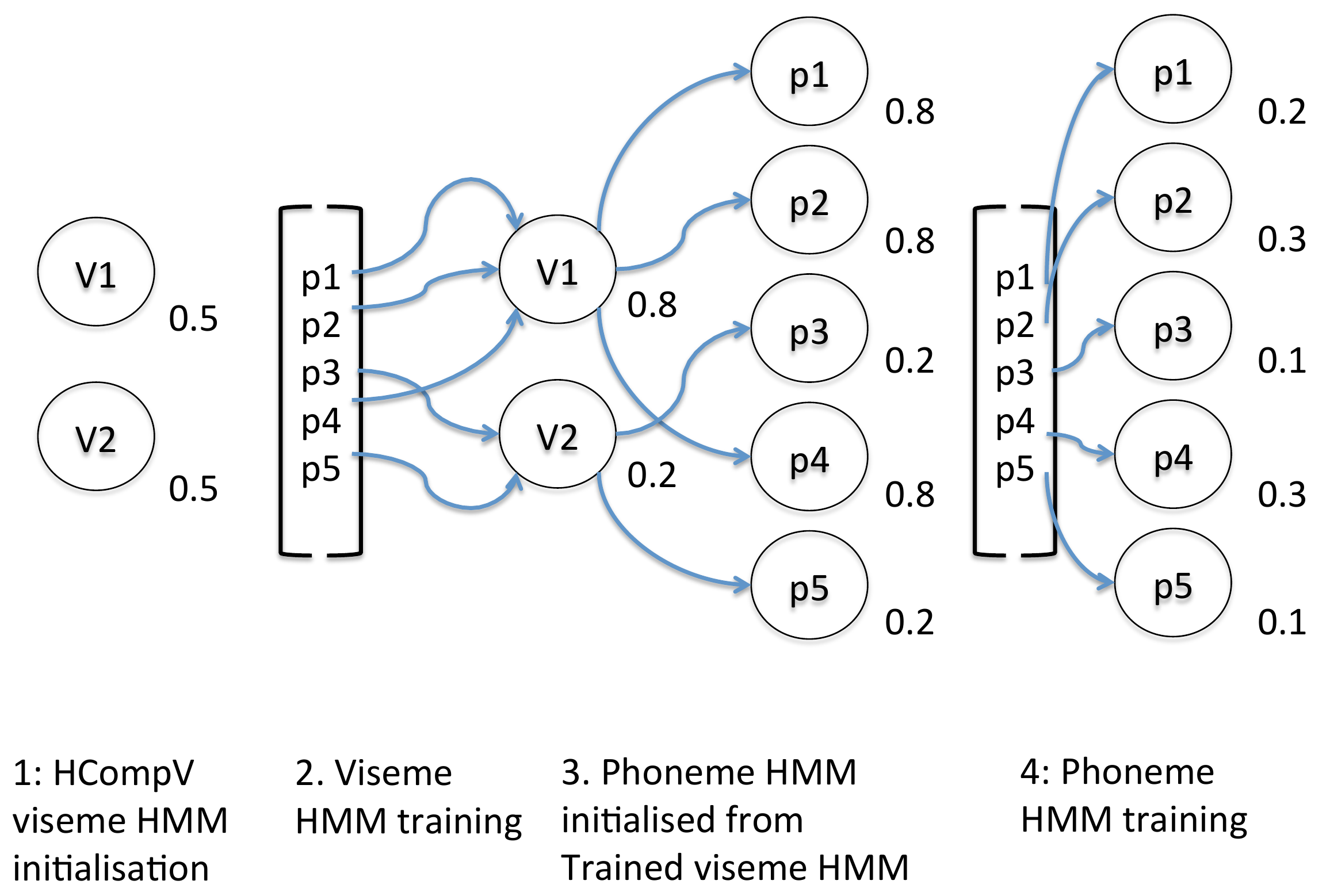}
\caption{Hierarchical training strategy for training visual units HHMs into phoneme-labeled HMM~classifiers.}
\label{fig:wlt_process}
\end{figure}

The basis of our new training algorithm is a hierarchical structure with the first level based on visual units, and~the second level based on phonemes. In~Figure~\ref{fig:wlt_process} we present an illustration based on a simple example using five phonemes (in reality there are up to $45$ in the RMAV sentences) mapped to two visual units (in reality there will be between $11$ and $35$ as we have refined our experiment to only use sets of visual units in the optimal size range from the preliminary test results). Each phoneme is mapped to a visual unit as in~\cite{bear2016decoding}, our example map is in Table~\ref{tab:exampleP2V}. But~now we are going to learn intermediate visual unit labeled HMMs before we create phoneme~models. 
In this example $/p1/$, $/p2/$ and $/p4/$ are associated with $/v1/$, so are initialized as duplicate copies of HMM $/v1/$. Likewise, phoneme models labeled $/p3/$ and $/p5/$ are initialized as replicas of $/v2/$. We now retrain the phoneme models using the same training~data.

\begin{table}[H]
\centering
\tablesize{\small} 

\caption{Our example P2V map to illustrate our novel training~algorithm}
\begin{tabular}{ll}
\toprule
\textbf{Visual Units} & \textbf{Phonemes} \\
\midrule
/v1/ & /p1/ /p2/ /p4/ \\
/v2/ & /p3/ /p5/ \\
\bottomrule
\end{tabular}
\label{tab:exampleP2V}
\end{table}
In full for each set of visual units of sizes from $11$ to $35$:
\begin{enumerate}[leftmargin=*,labelsep=4.9mm]
\item \textls[-15]{We initialize \textit{visual unit} HMMs with \texttt{HCompV}, this tool initializes HMMs defines all models equal~\cite{young2006htk}.} 
\item With our prototype HMM based upon a Gaussian mixture of five components and three states, we use \texttt{HERest} 11 times over to re-estimate the HMM parameters and we include short-pause model state tying (between re-estimates three and four with \texttt{HHed}). Training samples are from all phonemes in each visual unit cluster. These first two points are steps 1 and 2 in Figure~\ref{fig:wlt_process}. 
\item \textls[-15]{Before classification, our visual unit HMM definitions duplicated to be used as initialized definitions} for phoneme-labeled HMMs (Figure~\ref{fig:wlt_process} step 3). In~our Figure~\ref{fig:wlt_process} illustration, $/v1/$ is duplicated three times (one for each phoneme in its cluster) and $/v2/$ is copied twice. The~respective visual unit HMM definition is used for all the phonemes in its relative P2V map. 
\item These phoneme HMMs are retrained with \texttt{HERest} $11$ times over, this time, training samples are divided by the unique phoneme labels. 
\item We create a bigram word lattice with \texttt{HLStats} and \texttt{HBuild} and as part of the classification we apply a grammar scale factor of $1.0$ and a transition penalty of $0.5$ (based on~\cite{howell2013confusion}) with \texttt{HVite}. In~Section~\ref{sec:ln} we present a test to determine the best language network units for this step. 
\item Finally, the~output transcripts from \texttt{HVite} are used in \texttt{HResults} against the phoneme ground truths produced by \texttt{HLed}. This is all implemented using $10$-fold cross-validation with replacement~\cite{efron1983leisurely}. 
\end{enumerate}

The big advantage of this approach is the phoneme classifiers have seen mostly positive cases therefore have good mode matching, the~disadvantage is they are limited in their exposure to negative cases, less so than the visual~units. 

\section {Language Network Units}
\label{sec:ln}
Step five in our novel hierarchical training method requires a language network. It has been consistently observed that language models are very powerful in lipreading systems (e.g. in~\cite{6288999}). Language models built upon the ground truth utterances of datasets learn grammar and structure rules of words and sentences (the latter in the case of continuous speech). However, the~visual co-articulation effects damages the performance of visual speech language models as visually, people do not say what the language model expects. These types of network are commonplace, but we note that higher-order $N$-gram language models may improve classification rates but the cost of this model is disproportionate to our goal of developing more accurate classifiers. Therefore, to~decide which unit would best optimize our language model we test three units: visemes; phonemes; and words, as~bigram models in a second preliminary~test.

In the first two columns of Table~\ref{tab:sn_tests2} we list the possible pairs of classifier units and language model units. For~each of these pairs we use the common process previously described for lipreading in HTK, where our phonemes are based on the International Phonetic Alphabet~\cite{international1999handbook}, and~our visemes are Bear's speaker-dependent visemes~\cite{bear2017phoneme}. Word labels are from the RMAV dataset. We define \textbf{classifier units} as the labels used to identify individual classification models and \textbf{language units} as the label scheme used for building the decoding network used post~classification. 

\subsection{Language Network Unit~Analysis}
\begin{figure}[H]
  \centering
\begin{tabular}{c}
\includegraphics[width=0.65\textwidth]{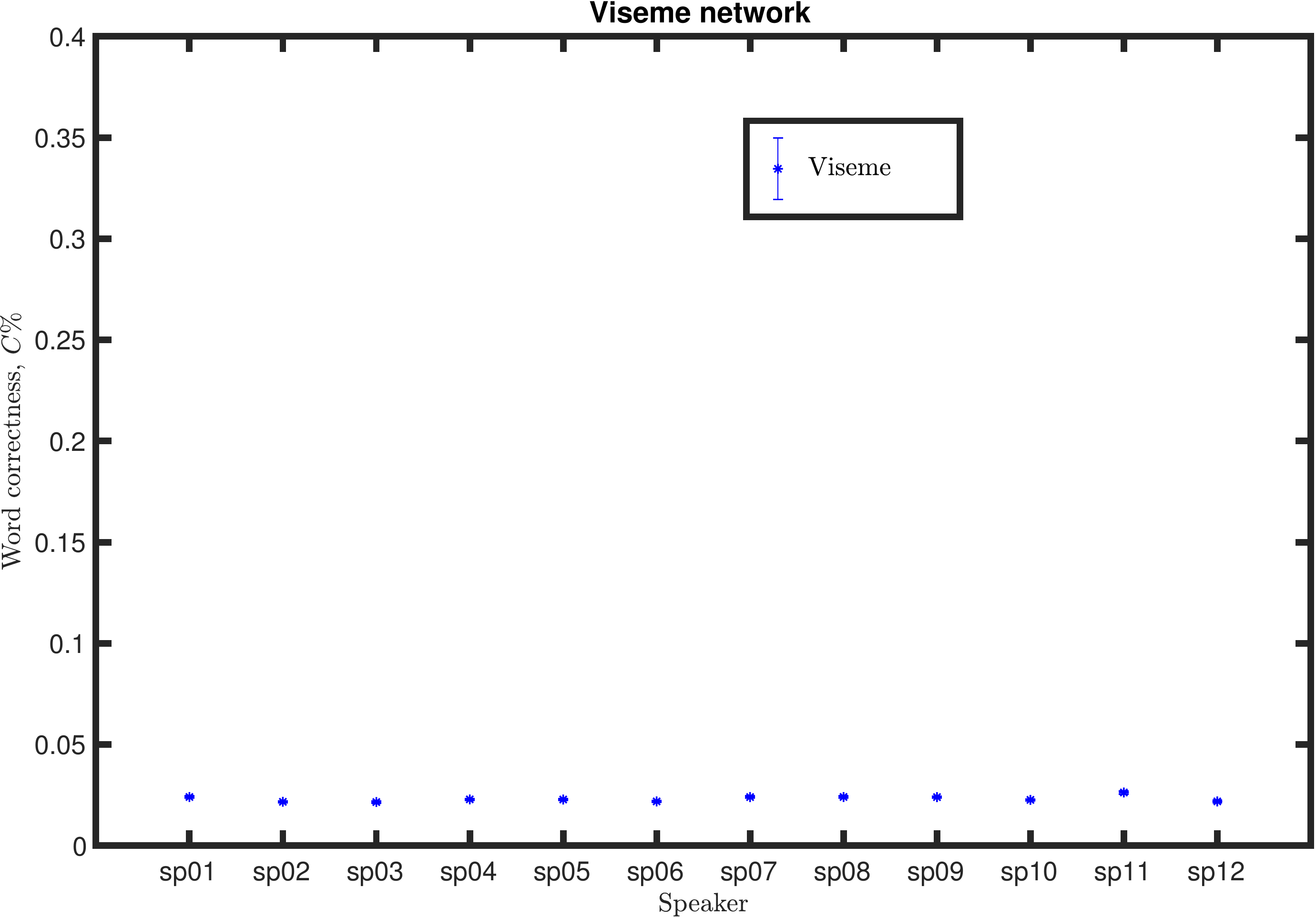} \\
\includegraphics[width=0.65\textwidth]{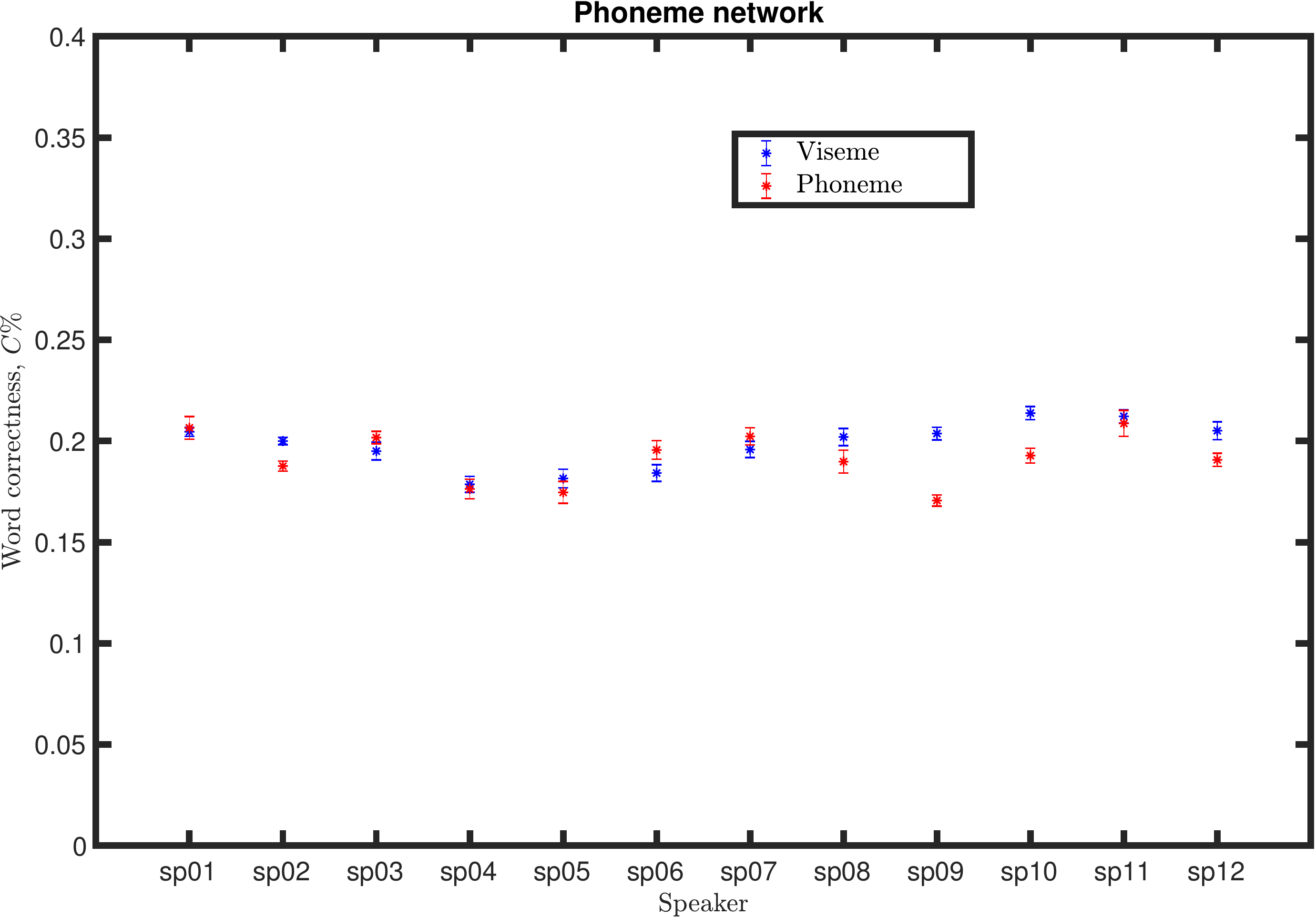} \\
\includegraphics[width=0.65\textwidth]{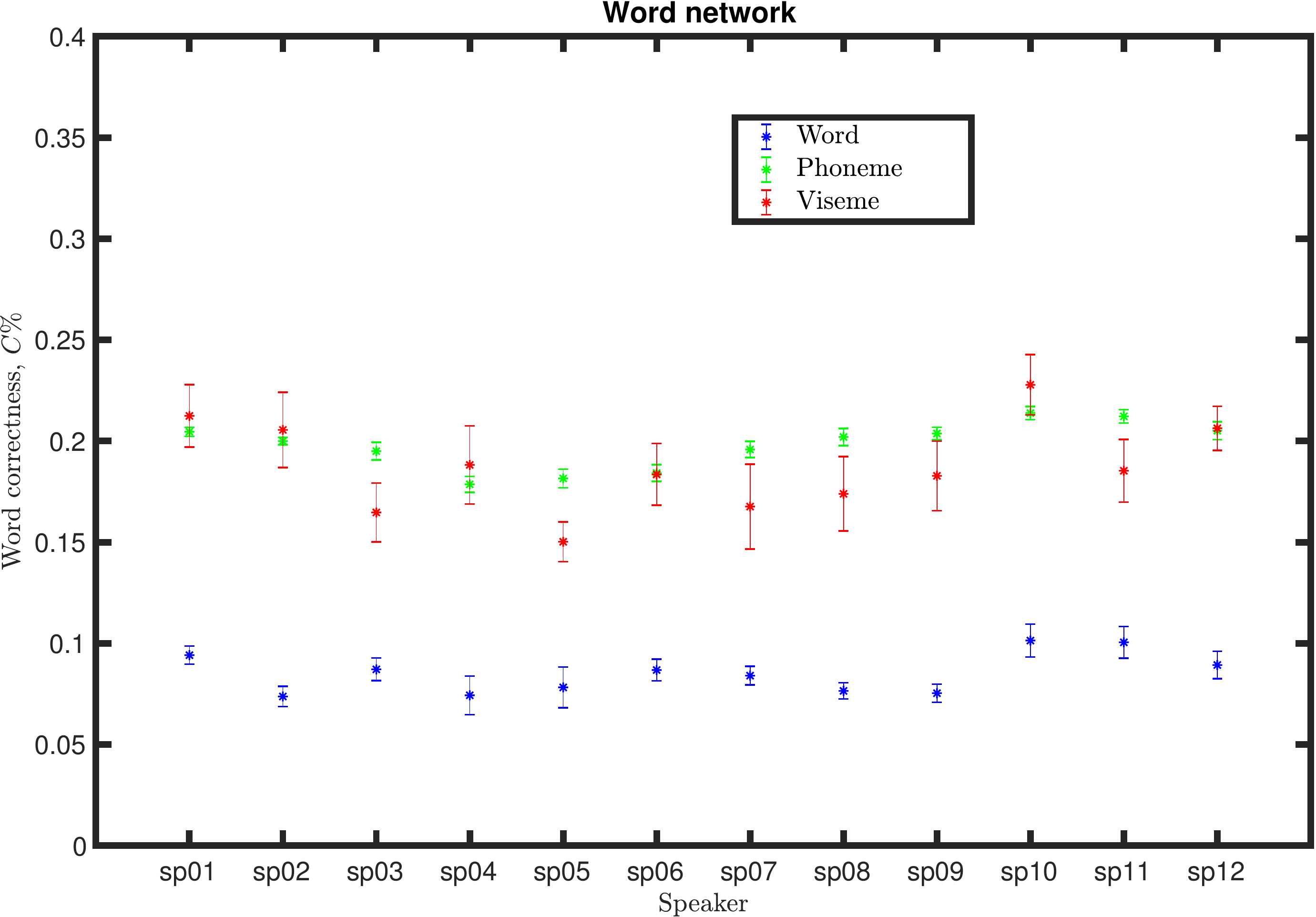} 
\end{tabular}
\caption{Effects of support network unit choice with each type of labeled HMM classifier units. Along~the $x$-axis is each speaker, $y$-axis values are correctness, $C$. Viseme network is at the top, phoneme network plotted in the middle, and~word networks at the~bottom.}
\label{fig:sn_effects_talker}
\end{figure}

In Table~\ref{tab:sn_tests2} column four we have listed one standard error values for these tests. The phoneme units are the most robust. 
In Figure~\ref{fig:sn_effects_talker} we have plotted word correctness ($x$-axis) for each speaker along the $y$-axis over three figures, one figure per language network unit. The~viseme network is top, phoneme network middle, and~word network at the bottom. The~viseme network is the lowest performing score ($0.02\pm0.0063$). On~the face of it, the~idea of visemes classifiers is a good one because they take visual co-articulation into account to some extent. However, as~seen here, a~language model of visemes is too complex because of homophenes. This leaves us with a choice of either phoneme or word units for our language model in step five of our new hierarchical training~method. 

\begin{table}[H]
\centering
\caption{Unit selection pairs for HMMs and language network combinations, and~the all-speaker mean $C_w$ achieved.} 
\begin{tabular}{llrr}
\toprule
\textbf{Classifier units} & \textbf{Network units} & \boldmath{$C_w$} & \textbf{$1$se} \\
\midrule
Viseme & Viseme & $0.02$ & $0.0063$		\\
\midrule
Viseme & Phoneme & $0.19$ & $0.0036$ 		 \\
Phoneme & Phoneme & $0.19$ & 	$0.0036$ \\
\midrule
Viseme & Word & $0.09$ &  $0.0$		 \\
Phoneme & Word & $0.20$ & 	$0.0043$	\\
Word & Word & $0.19$ & $0.0005$\\
\bottomrule
\end{tabular}
\label{tab:sn_tests2}
\end{table}

In Figure~\ref{fig:sn_effects_talker} (middle) we have our phoneme language network performance with both viseme and phoneme trained classifiers. This is more exciting because for all speakers we see a statistically significant increase in $C_w$ compared to the viseme network scores in Figure~\ref{fig:sn_effects_talker} top. Looking more closely between speakers we see that for four speakers (2, 9, 10 and 12), the~viseme classifiers outperform the phonemes, yet for all other speakers there is no significant difference between the two. On~average they are identical with an all-speaker mean $C_w$ of $0.19\pm0.0036$ compared to the viseme classifiers (Table~\ref{tab:sn_tests2}, column 3). 

In Figure~\ref{fig:sn_effects_talker} (bottom) we show our $C_w$ for all speakers with a word network paired with classifiers built on viseme, phoneme, and~word units. Our first observation is that word classifiers perform very poorly. We attribute this to a low number of training samples per class due to the extra number of classes in the word space compared to the number of classes in the phoneme space, so we do not continue our work with word-based classifiers.
Also shown in Figure~\ref{fig:sn_effects_talker} (bottom) are the phoneme and viseme classifiers (in green and red respectively) with a word network. This time we see that for five of our 12 speakers (3, 5, 7, 8, and~11), the~phoneme classifiers outperform the visemes and for our remaining speakers there is no significant difference once a work network is~applied. 

These results tell us that for some speakers viseme classifiers with phoneme networks are a better choice whereas others are easier to lipread with phoneme classifiers with a word network. Thus, we continue our work using both phoneme and word-based language~networks. 

\section{Effects of Training Visual Units for Phoneme~Classifiers}
\begin{figure}[H]
\centering
\includegraphics[width=0.9\textwidth]{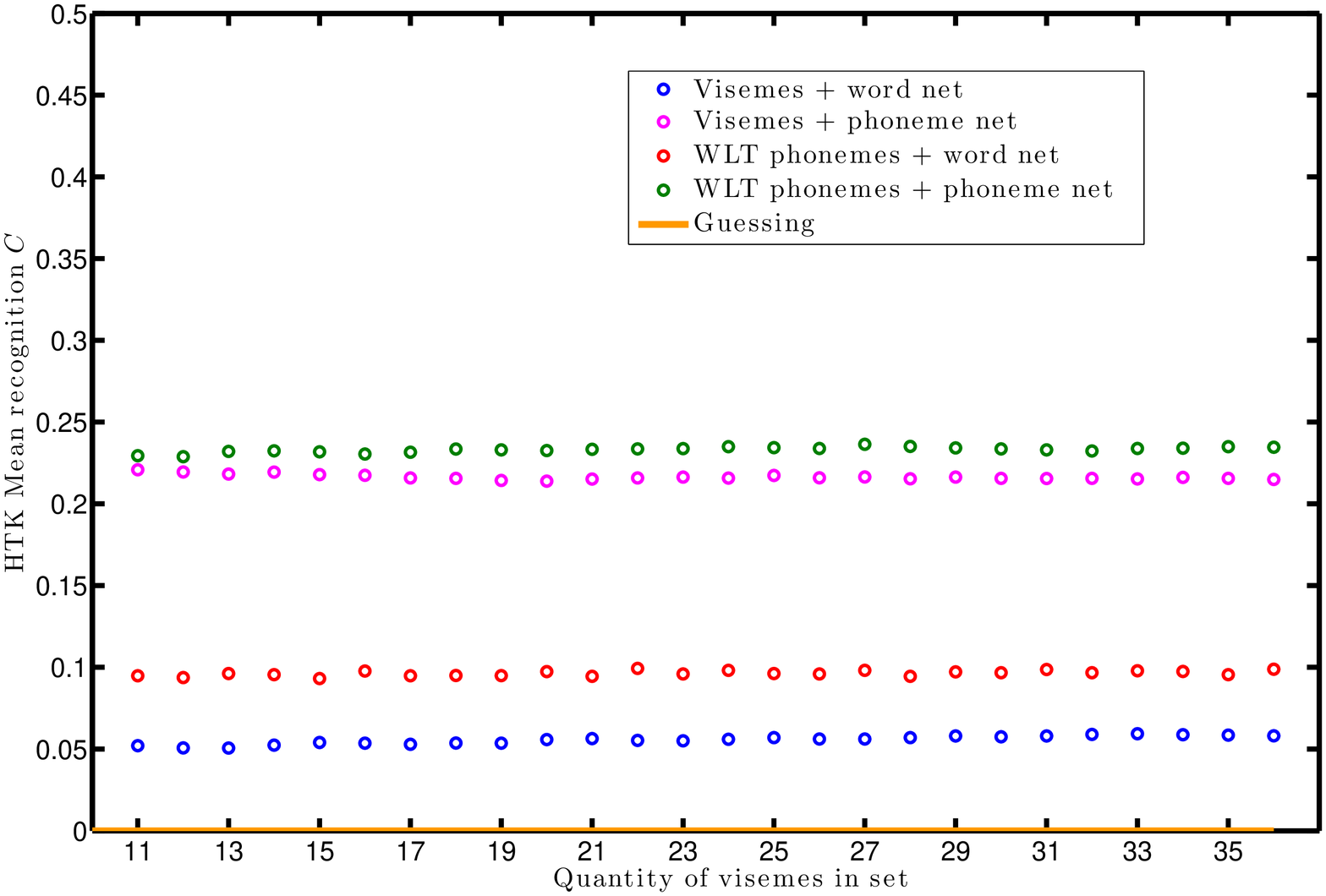}
\caption{HTK Correctness $C$ for visual unit classifiers with either phoneme or word language models and phoneme classifiers with either phoneme or word language models averaged over all 12 speakers. The~correctness unit matches the paired network~unit. }
\label{fig:res_all}
\end{figure}

Here we present the results of our proposed hierarchical training method (described in Section~\ref{sec:newclusteringalg} with two different language models. Figure~\ref{fig:res_all} shows the mean correctness, $C$, for~all 12 speakers over $10$ folds. We have plotted four symbols, one for each of the pairings of our HMM unit labels and the language network unit (\{visual units and phonemes, visual units and words, phonemes and phonemes, phonemes and words\}). Random guessing is plotted in~orange.

\begin{table}[H]
\centering
\caption{Minimum and maximum all-speaker mean correctness, $C$, showing the effect of hierarchical training from visual units on phoneme-labeled HMM~classification.}
\begin{tabular}{lrrr}
\toprule
 	& \textbf{Min} & \textbf{Max} & \textbf{Range} \\
\midrule
visual units + word net & 0.03 & 0.06 & 0.03 \\
Phonemes + word net & 0.09 & 0.10 & 0.01 \\
Effect of WLT & 0.06 & 0.04 & -- \\
visual units + phoneme net & 0.20 & 0.22 & 0.02 \\
Phonemes + phoneme net & 0.26 & 0.24 & 0.01 \\ 
Effect of WLT & 0.02 & 0.02 & -- \\
\bottomrule
\end{tabular}
\label{tab:meanstats}
\end{table}
\unskip

The $x$-axis of Figure~\ref{fig:res_all} is the size of the optimal visual unit sets from Figure~\ref{fig:correctness}, from~$11$ to $36$. This is the range of optimal number of visual units where phoneme label classifiers do not improve classification. The~baseline of visual unit classification with a word network from~\cite{bear2015findingphonemes} is shown in blue and is not significantly different from conventionally learned phoneme \textls[-15]{classifiers. Based on our language network study in Section~\ref{sec:ln}, it is not a surprise to see just by using a phoneme network instead of a word network to support visual unit classification we significantly improve our mean correctness score for all visual unit set sizes for all speakers (shown in pink). We have plotted weighted guessing in~orange.}

More interesting to see is our new weakly trained phoneme HMMs are significantly better than the visual unit HMMs. In~the first part of our work here phoneme HMMs gave an all-speaker mean $C = 0.059$ and was not significantly different from the best visual units. Here, regardless of the size of the original visual unit set, $C$ is almost double. Weakly learned phoneme classifiers with a word network gain $0.031$ to $0.040$ in mean $C$, and~when these phoneme classifiers are supported with a phoneme network we see a correctness gain range from $0.17$ to $0.18$. These gains are supported by the all-speaker mean minimum and maximums listed in Table~\ref{tab:meanstats}. These gain scores are from over all the potential P2V mappings and show there is little difference in which P2V map is best for knowing which set of visual units to initialize our phoneme classifiers. All results are significantly better than~guessing. 

\begin{figure}[H]
\centering
\includegraphics[width=0.68\textwidth]{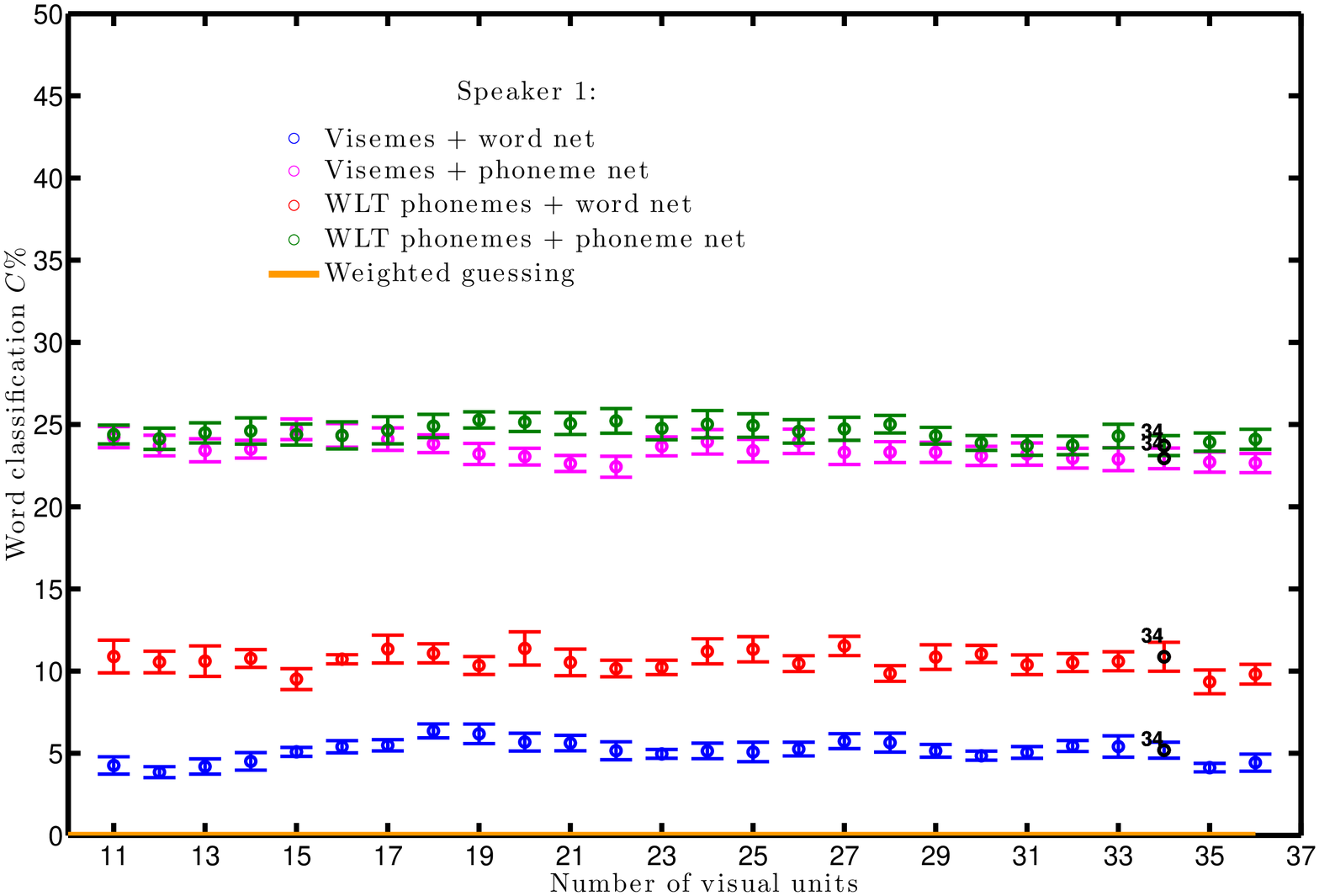} \\
\includegraphics[width=0.68\textwidth]{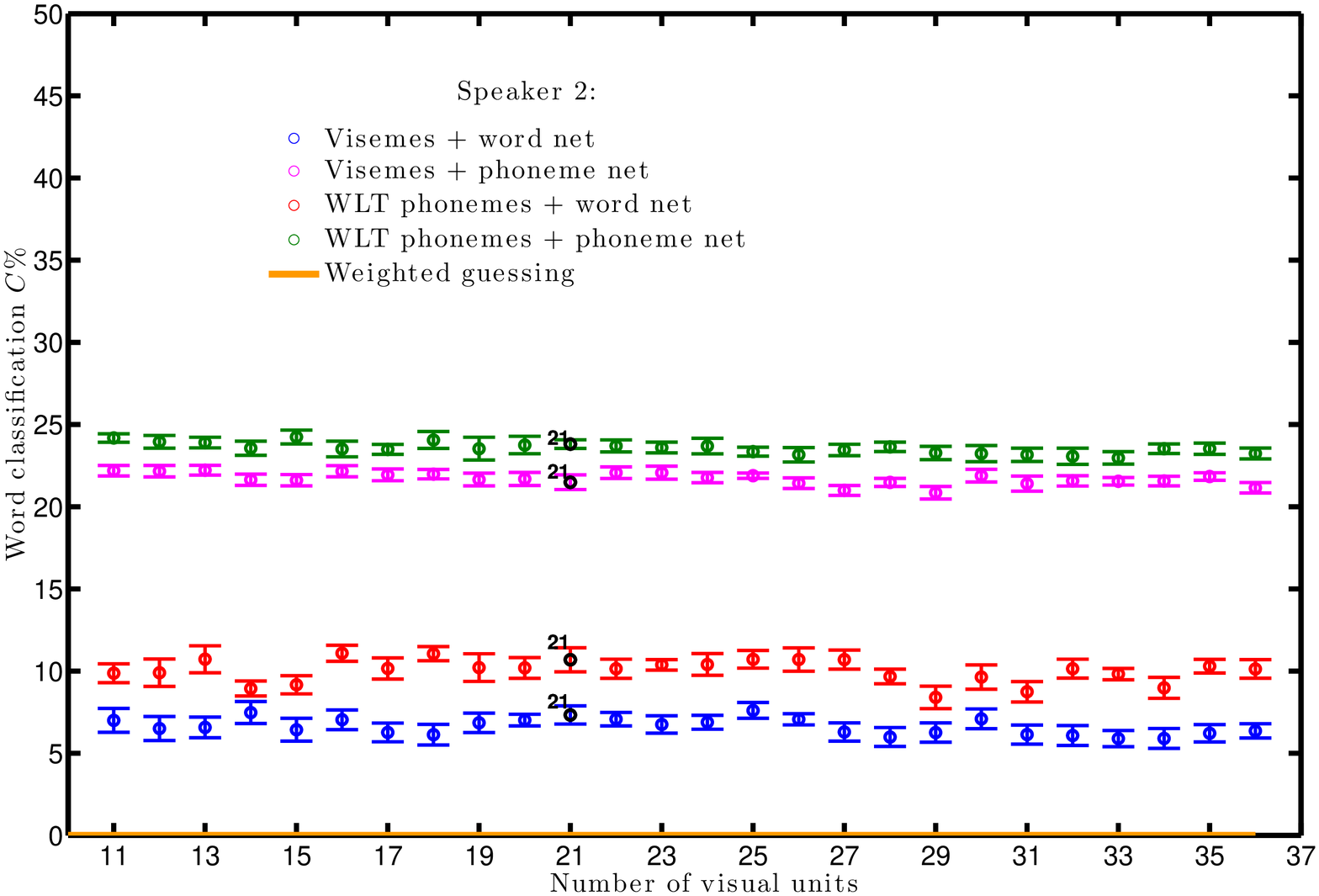} \\
\includegraphics[width=0.68\textwidth]{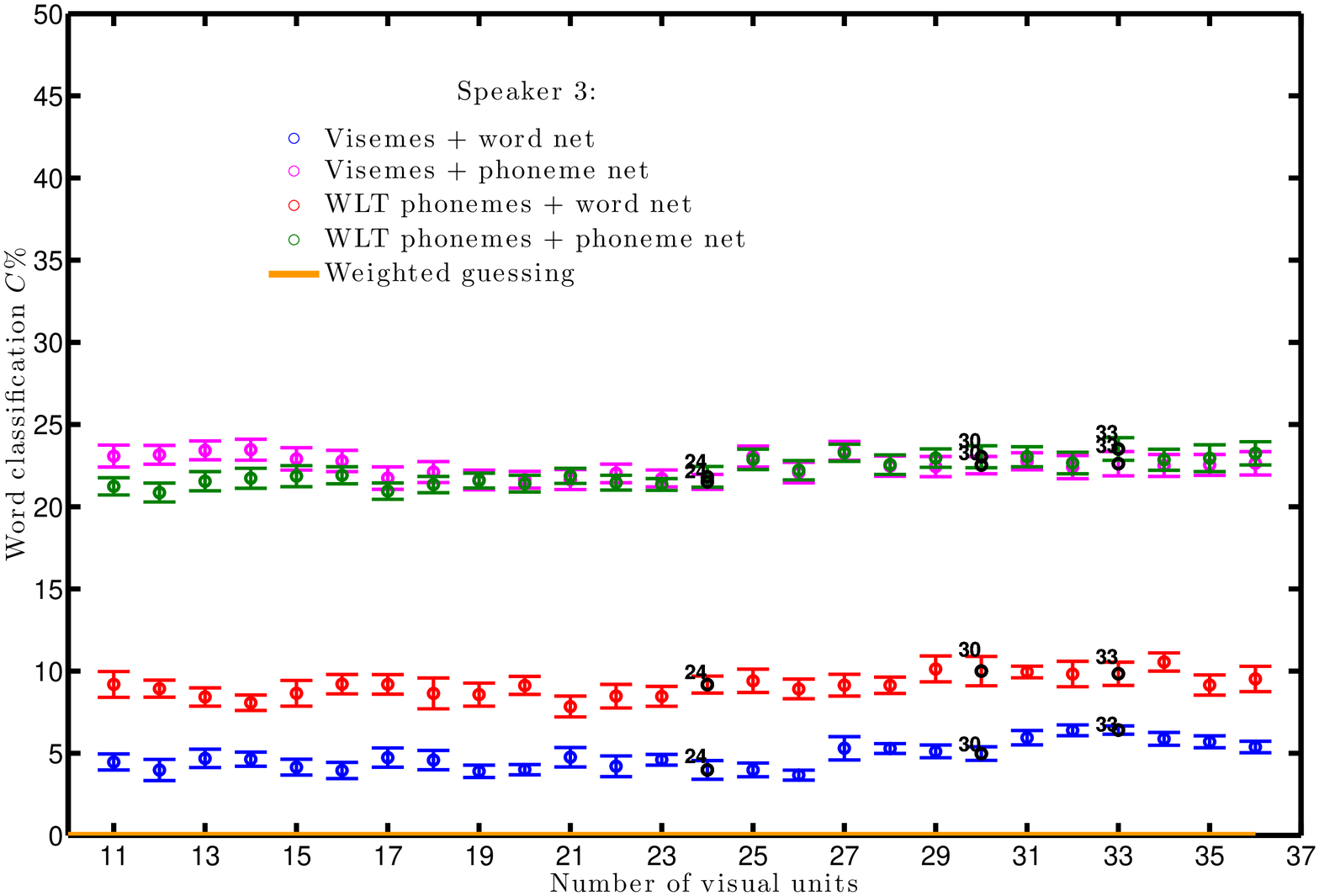} 
\caption{\textls[-15]{Speaker 1 (\textbf{top}), Speaker 2 (\textbf{middle}), and~Speaker 3 (\textbf{bottom}) correctness with a word language model (blue) and the hierarchically trained phoneme classifiers with a phoneme or word~network.}}
  \label{fig:rmav1wlt}
\end{figure}

\begin{figure}[H]
\centering
\includegraphics[width=0.68\textwidth]{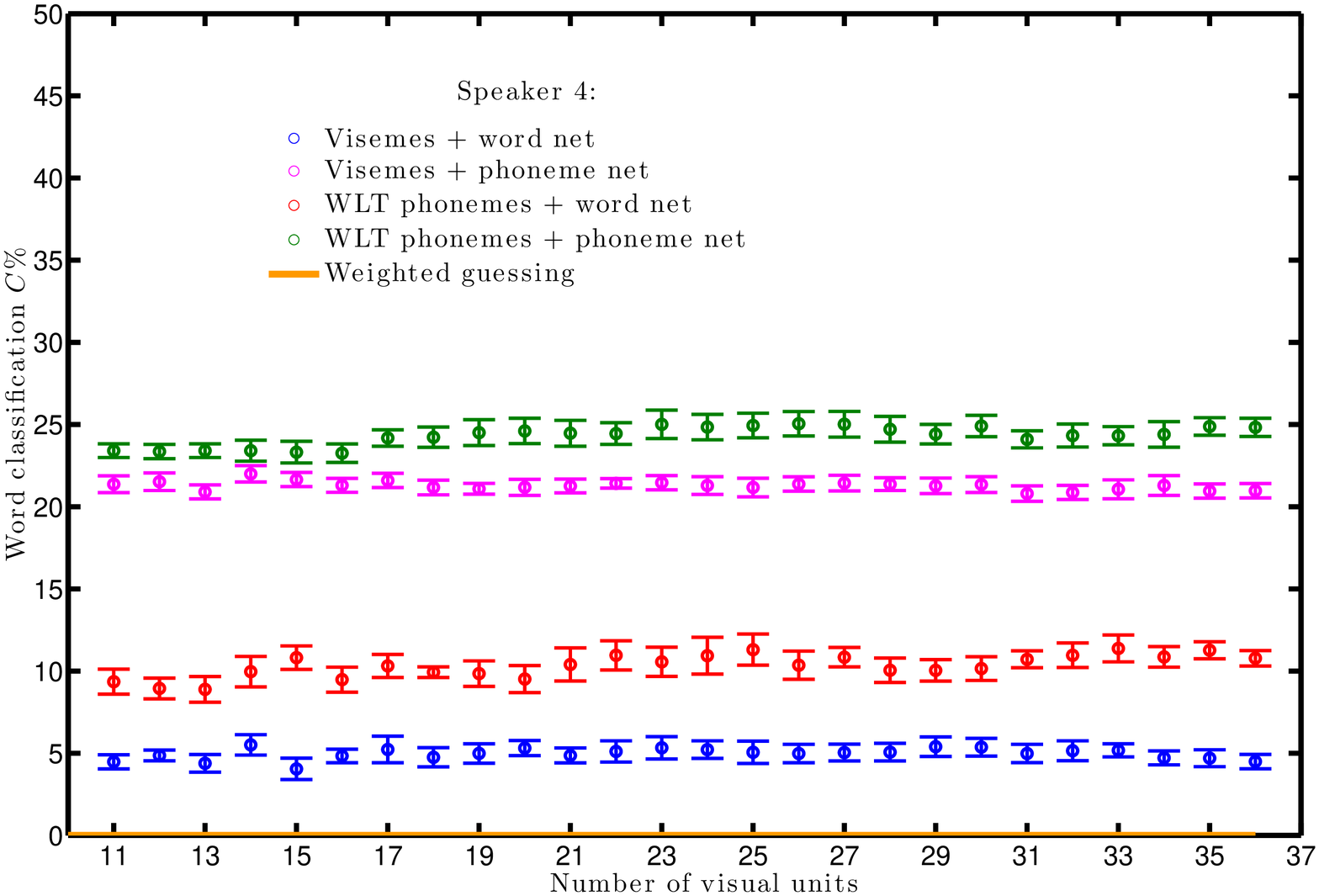} \\
\includegraphics[width=0.68\textwidth]{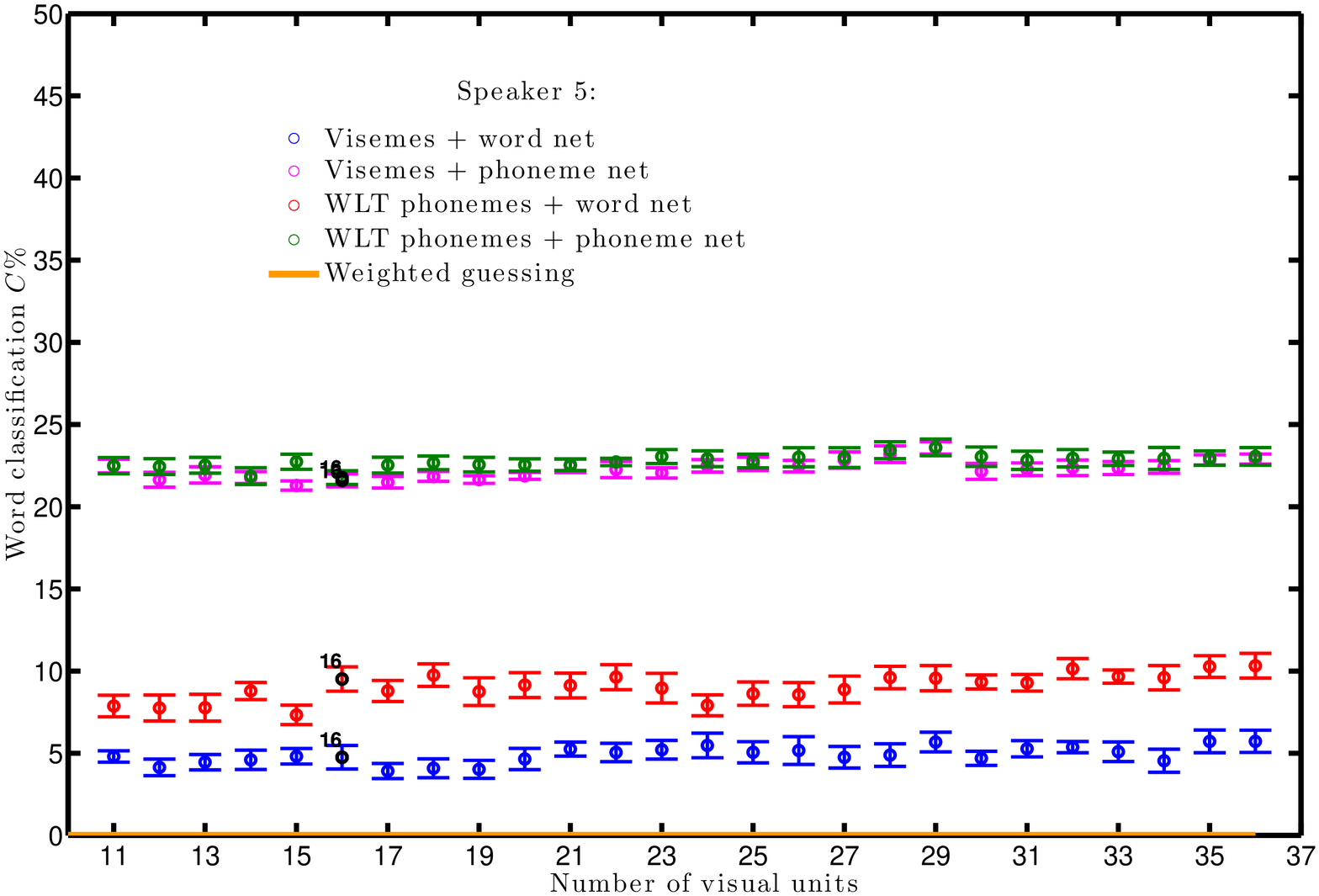} \\
\includegraphics[width=0.68\textwidth]{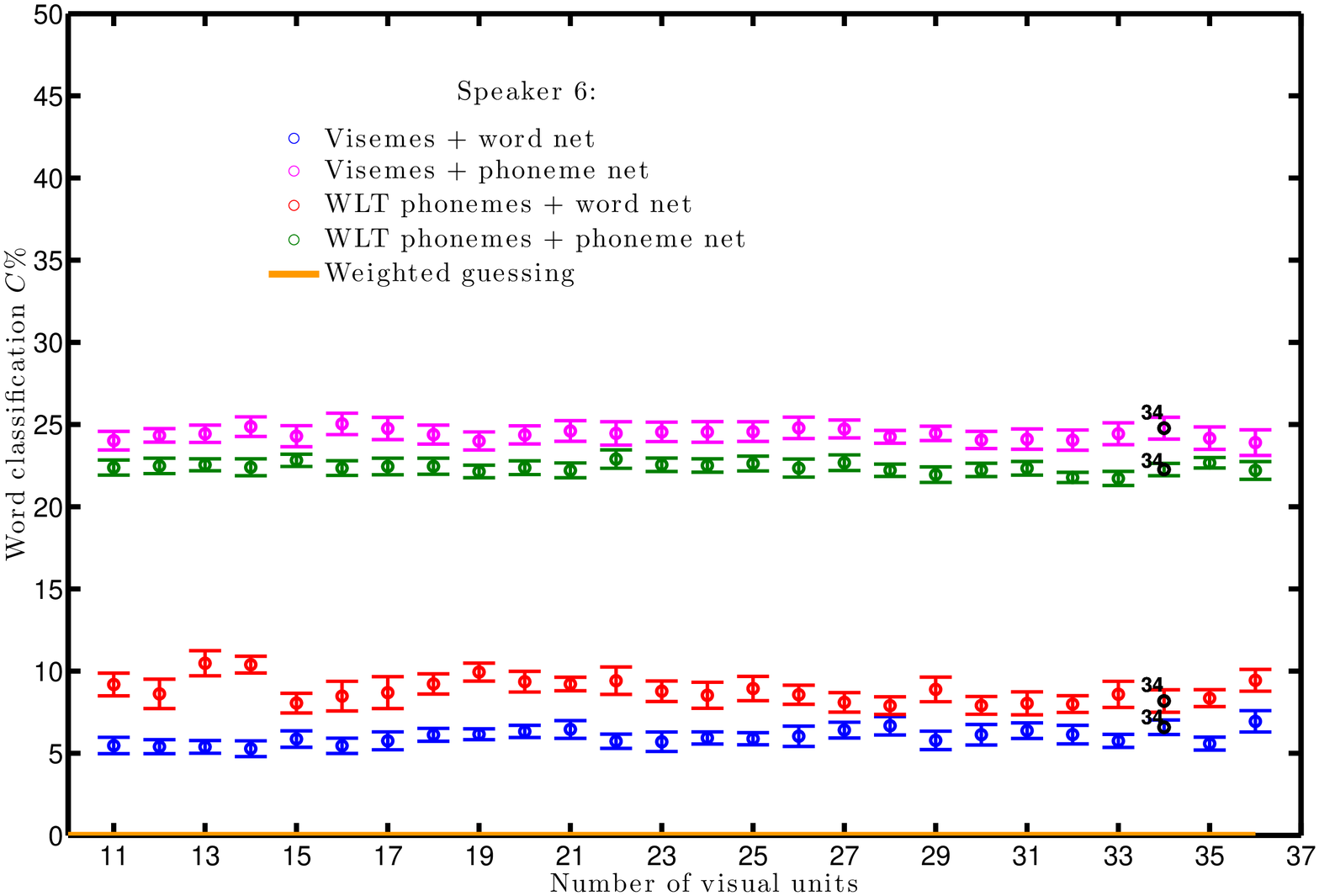} 
\caption{\textls[-15]{Speaker 4 (\textbf{top}), Speaker 5 (\textbf{middle}), and~Speaker 6 (\textbf{bottom}) correctness with a word language model (blue) and the hierarchically trained phoneme classifiers with a phoneme or word~network.}}
  \label{fig:rmav3wlt}
\end{figure}

\begin{figure}[H]
\centering
\includegraphics[width=0.68\textwidth]{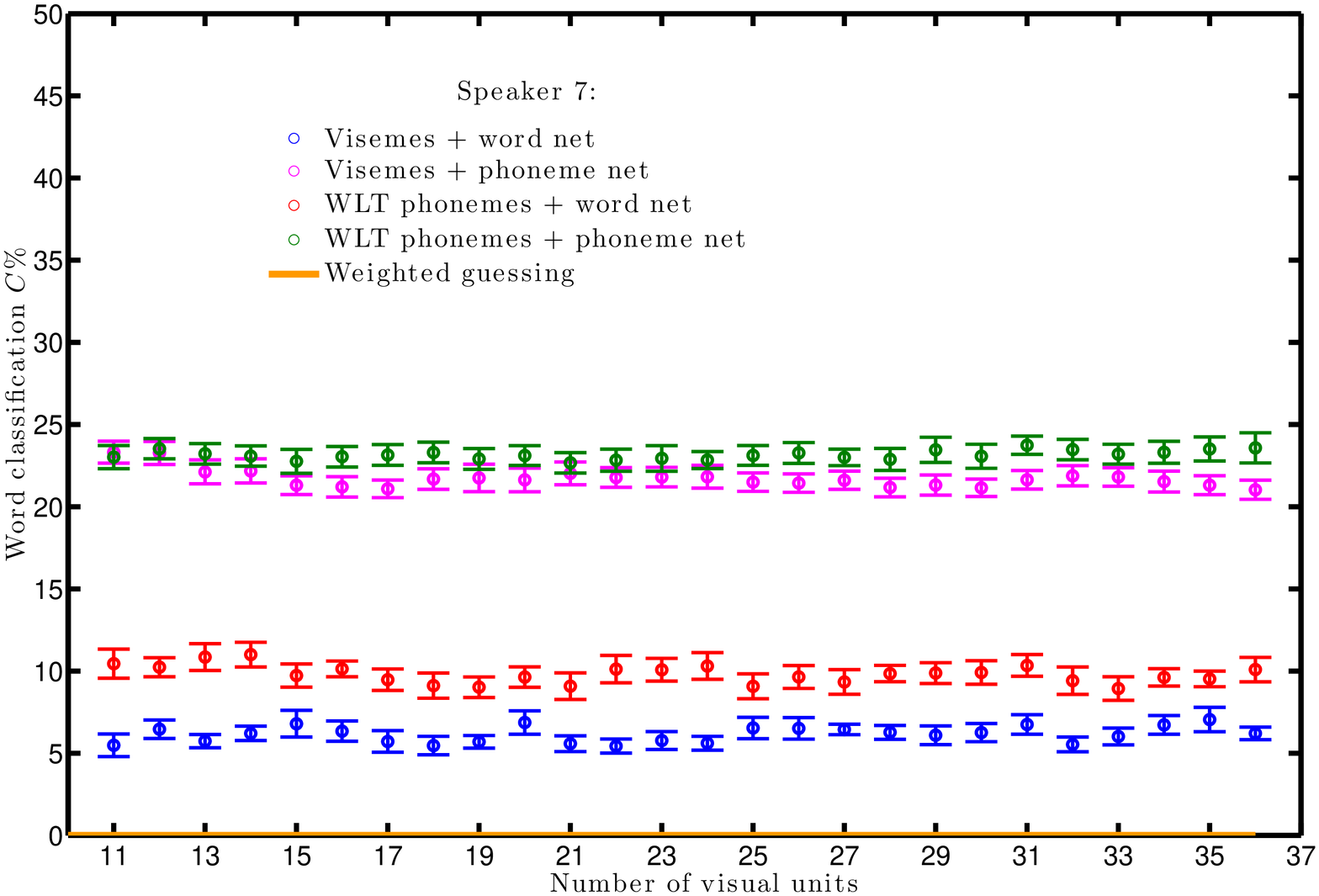} \\
\includegraphics[width=0.68\textwidth]{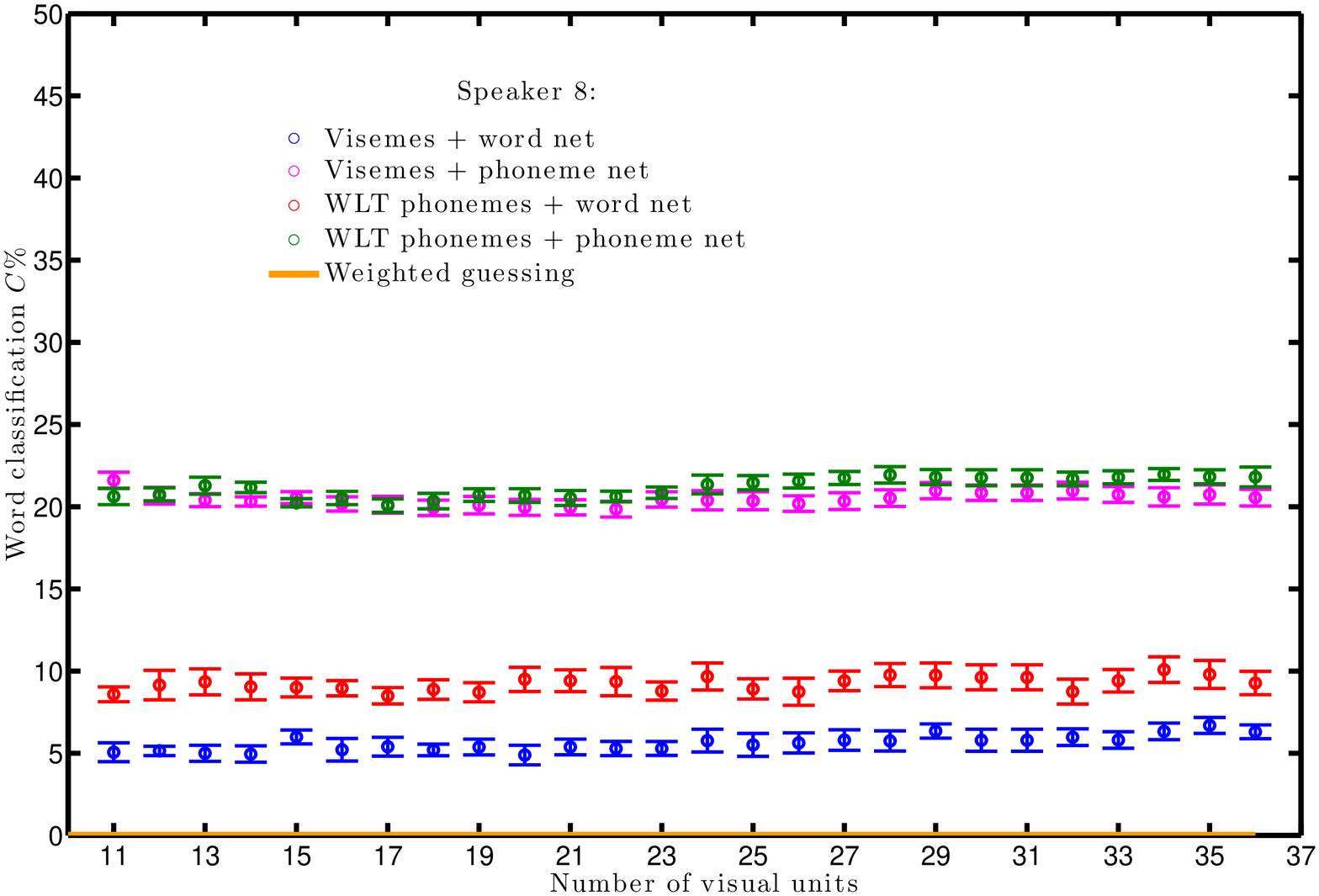} \\
\includegraphics[width=0.68\textwidth]{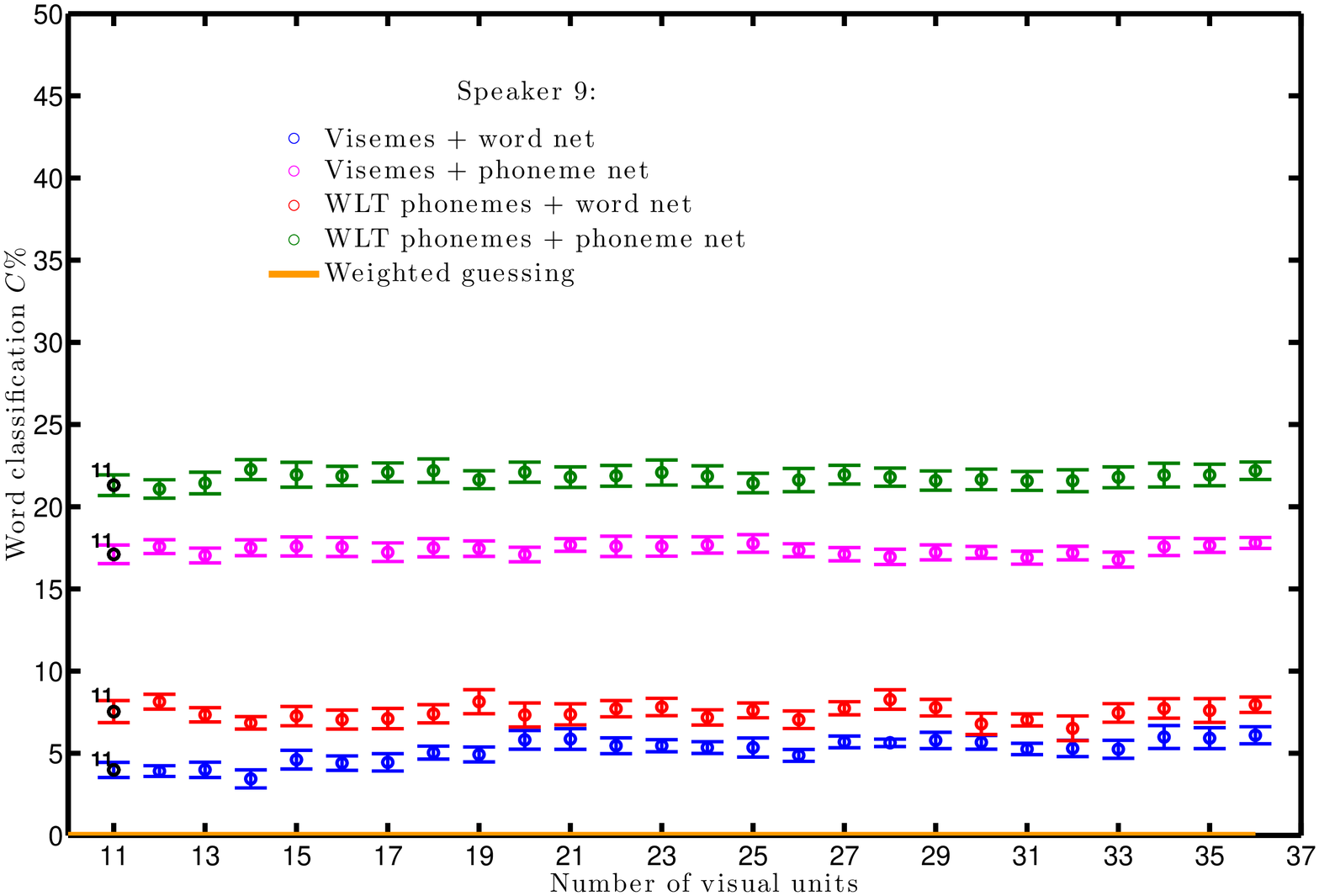} 
\caption{\textls[-15]{Speaker 7 (\textbf{top}), Speaker 8 (\textbf{middle}), and~Speaker 9 (\textbf{bottom}) correctness with a word language model (blue) and the hierarchically trained phoneme classifiers with a phoneme or word~network.}}
  \label{fig:rmav5wlt}
\end{figure}

\begin{figure}[H]
\centering
\includegraphics[width=0.68\textwidth]{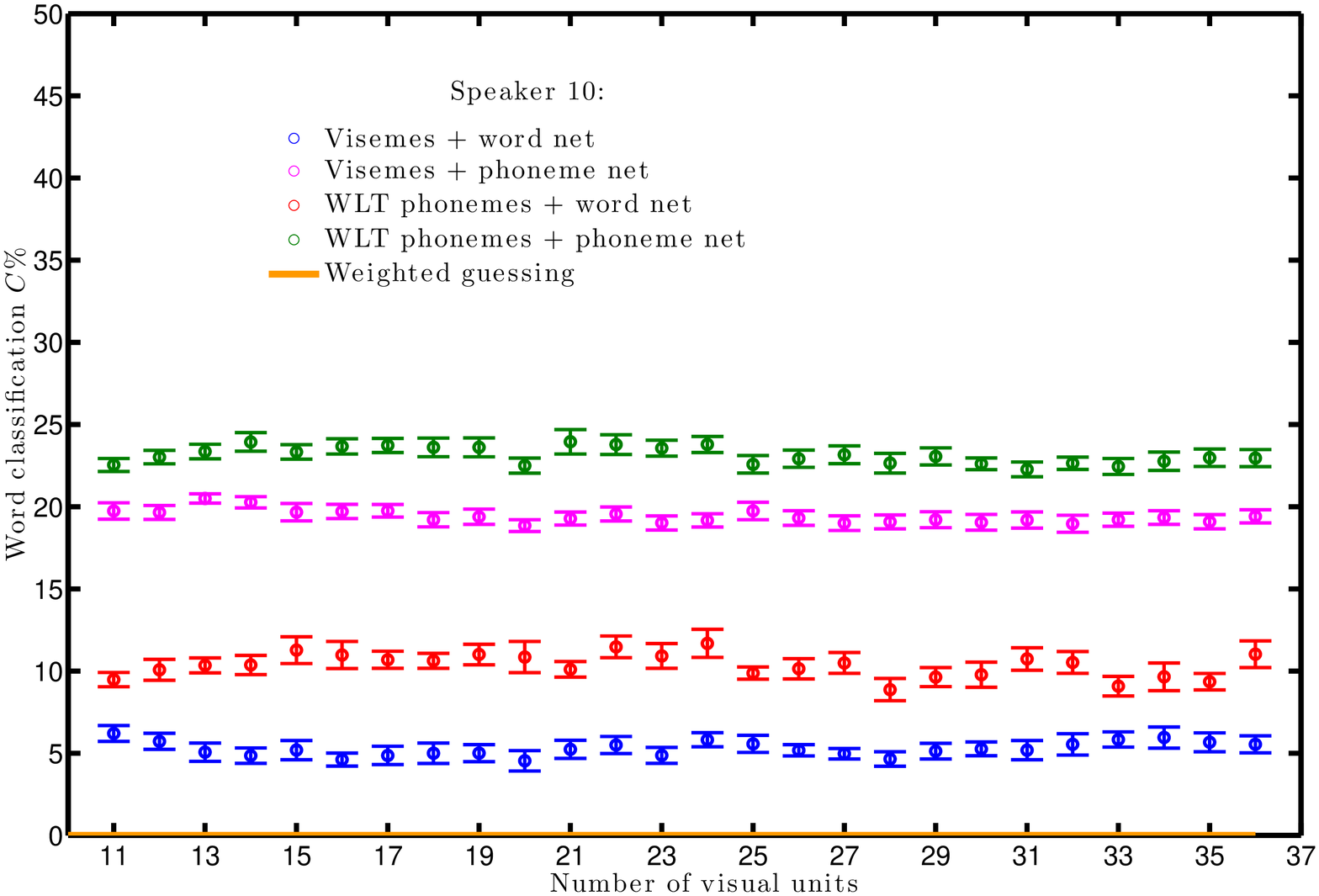} \\
\includegraphics[width=0.68\textwidth]{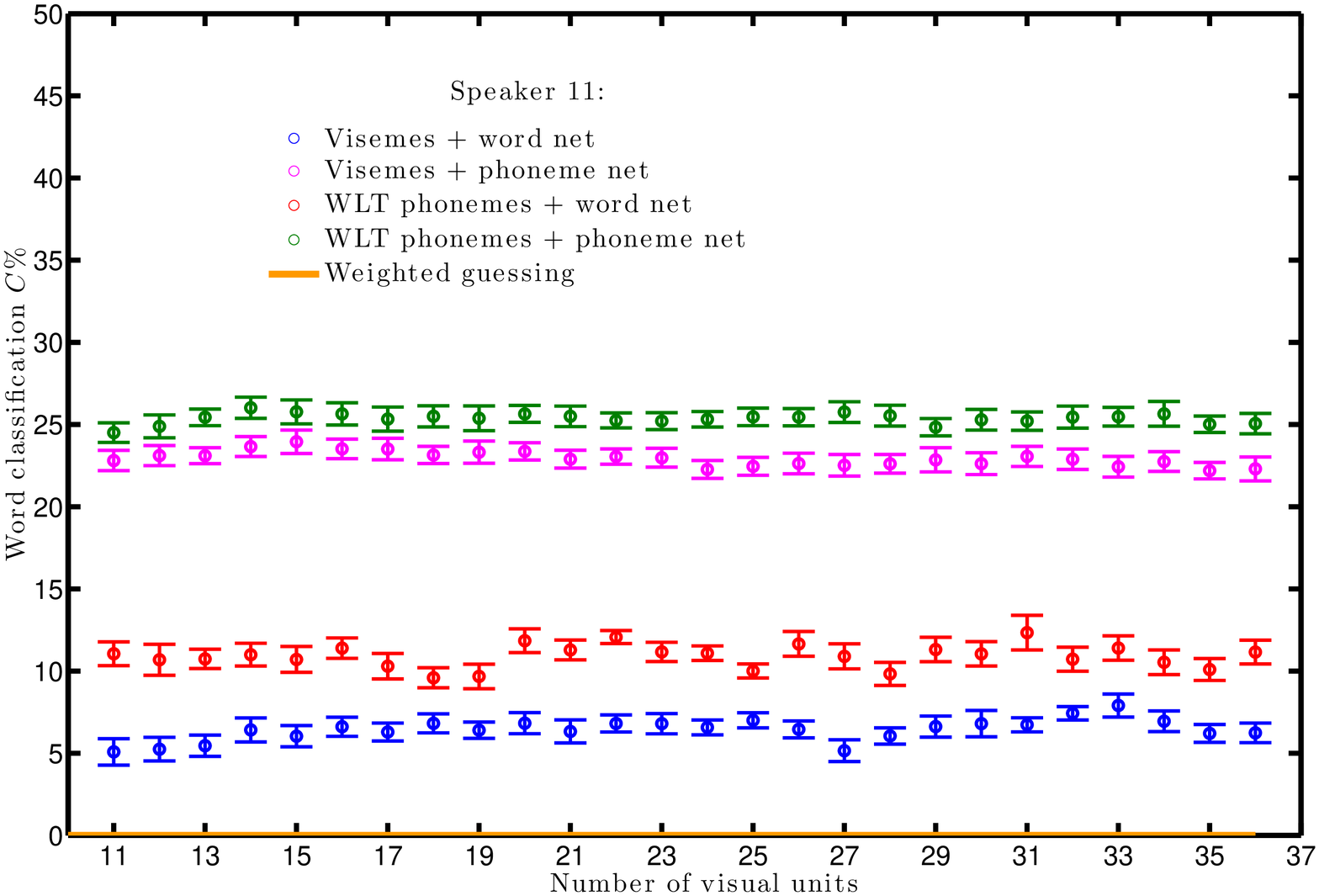} \\
\includegraphics[width=0.68\textwidth]{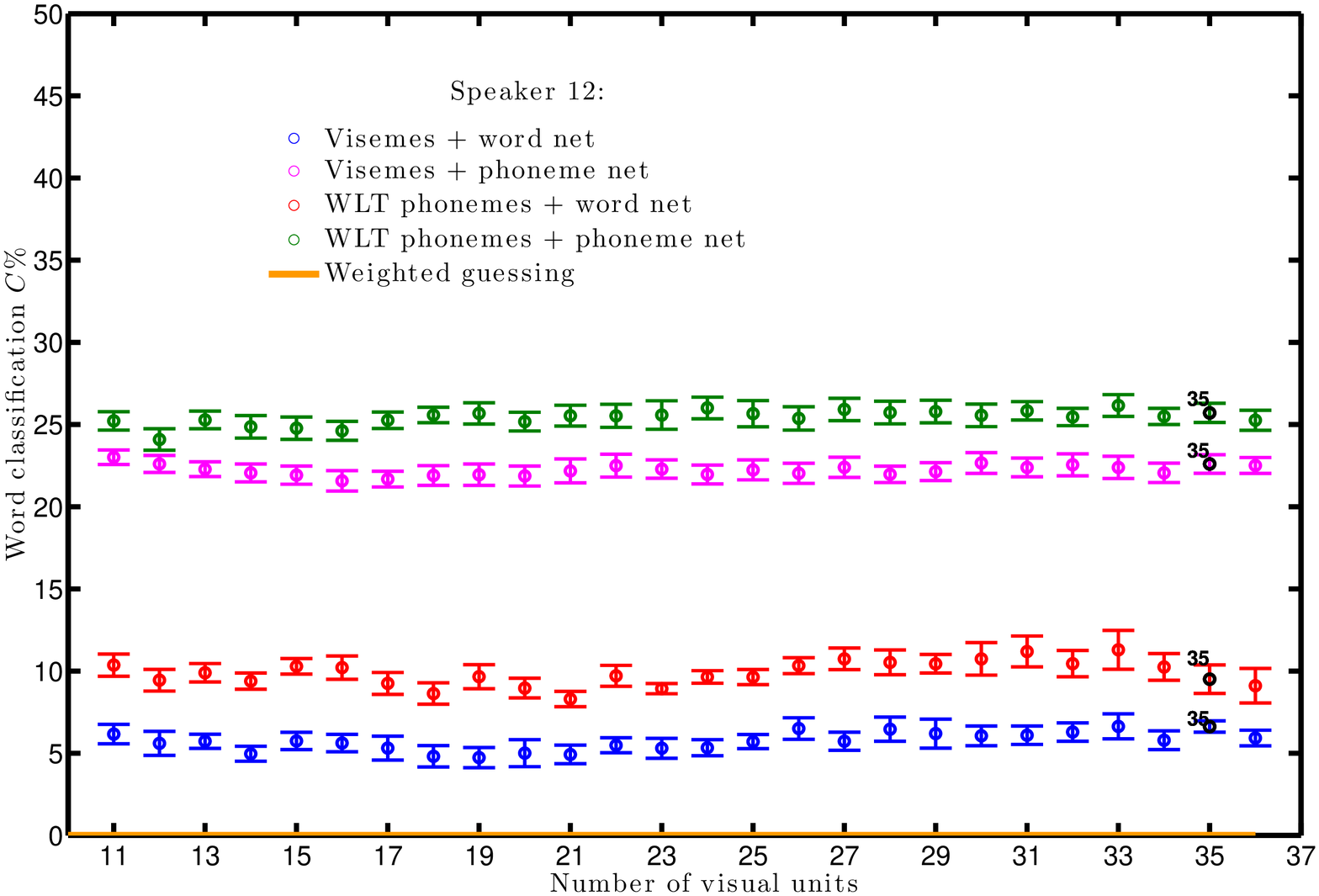} 
\caption{\textls[-15]{Speaker 10 (\textbf{top}), Speaker 11 (\textbf{middle}), and~Speaker 12 (\textbf{bottom}) correctness with a word language model (blue) and the hierarchically trained phoneme classifiers with a phoneme or word~network.}}
  \label{fig:rmav7wlt}
\end{figure}

In Figures~\ref{fig:rmav1wlt}--\ref{fig:rmav7wlt}, we have plotted for each of our $12$ speakers non-aggregated results showing $C \pm$ one standard error. While not monotonic, these graphs are much smoother than the speaker-dependent graphs shown in appendix A. The~significant differences between visual unit set sizes (in Figure~\ref{fig:correctness}) have now disappeared because the learning of differences between visual units, has been incorporated into the training of phoneme classifiers, which in turn are now better trained (plotted in red and green which improve on blue and pink respectively).

\unskip

\unskip

\unskip

An intriguing observation is comparing the use of a phoneme network for visual units and for weakly taught phonemes. For~some speakers, the~weakly learned phonemes are not always as important as having the right network unit. This is seen in Figure~\ref{fig:rmav1wlt} (top and bottom),~Figure \ref{fig:rmav3wlt} (middle),~Figure \ref{fig:rmav5wlt} (middle), and Figure \ref{fig:rmav7wlt} (bottom) for Speaker's 1, 3, 5, 8, and~12. By~rewatching the original videos to estimate the age of our speakers, we categorize them as either an `older' or 'younger' speaker by eye because the exact ages were not captured during filming. The~speakers with less significant difference in the effect of hierarchical training from visual to audio units are younger. This implies to lipread a younger person we need more support from the language model, than~an older speaker. We suggest this could be because young people show more co-articulation than older people, but~this requires further~investigation. 

\section{Conclusions}
We have described a method that allows us to construct any number of visual units. The~presence of an optimum is a result of two competing effects on a lipreading system. In~the first, as~the number of visual units shrinks the number of homophenes rises and it becomes more difficult to recognize words (correctness drops). In~the second, as~the number of visual units rises we run out of training data to learn the subtle differences in lip-shapes (if they exist), so again, correctness drops. Thus, the~optimum number of visual units lies between one and $45$. In~practice we see this optimum is between the number of phonemes and eight (which is the size of one of the smaller visual unit sets). 

The choice of visual units in lipreading has caused some debate. Some workers use visemes (for~example Fisher~\cite{fisher1968confusions} in which visemes are a theoretical construct representing phonemes that should look identical on the lips~\cite{hazen2006visual}). Others, e.g. ~\cite{howell2013confusion} have noted that lipreading using phonemes can give superior performance to visemes. 
Here, we supply further evidence to the more nuanced hypothesis first presented in~\cite{bear2015findingphonemes}, that there are intermediate units, which for convenience we call visual units, that can provide superior performance provided they are derived by an analysis of the data. A~good number of visual units in a set is higher than previously~thought. 




We have also presented a novel learning algorithm which shows improved performance for these new data-driven visual units by using them as an intermediate step in training phoneme classifiers. The~essence of our method is to retrain the visual unit models in a fashion similar to hierarchical training. This two-pass approach on the same training data has improved the training of phoneme-labeled classifiers and increased the classification~performance. 

We have also investigated the relationship between classifier unit choice with the unit choice for the supporting language network. We have shown that one can choose either phoneme or words without significantly different accuracy, but~recommend a word net as this reduces the effect of homophene error and enables unbiased comparison of classifier~performance. 

In future works we would seek to experiment if this hierarchical training method would achieve the same benefit to other classification techniques, for~example RBMs. This is inspired by the work in~\cite{mousas2017real,nam2012learning} and other recent hybrid HMM studies such as~\cite{thangthai2018computer}.

\vspace{6pt}

\authorcontributions{H.B. and R.H conceived and designed the experiments; H.B.performed the experiments; H.B and R.H analyzed the data; H.B and R.H wrote the paper.} 

\funding{This research was funded by EPSRC grant number 1161995. The APC was funded by Queen Mary University of London.}

\acknowledgments{We would like to thank our colleagues at the University of Surrey who were collaborators in building the RMAV dataset.}
\conflictsofinterest{The authors declare no conflict of interest.}


\appendixtitles{no} 
\appendix
\section{}
\label{sec:app}
\begin{figure}[H] 
\centering 
\includegraphics[width=0.49\textwidth]{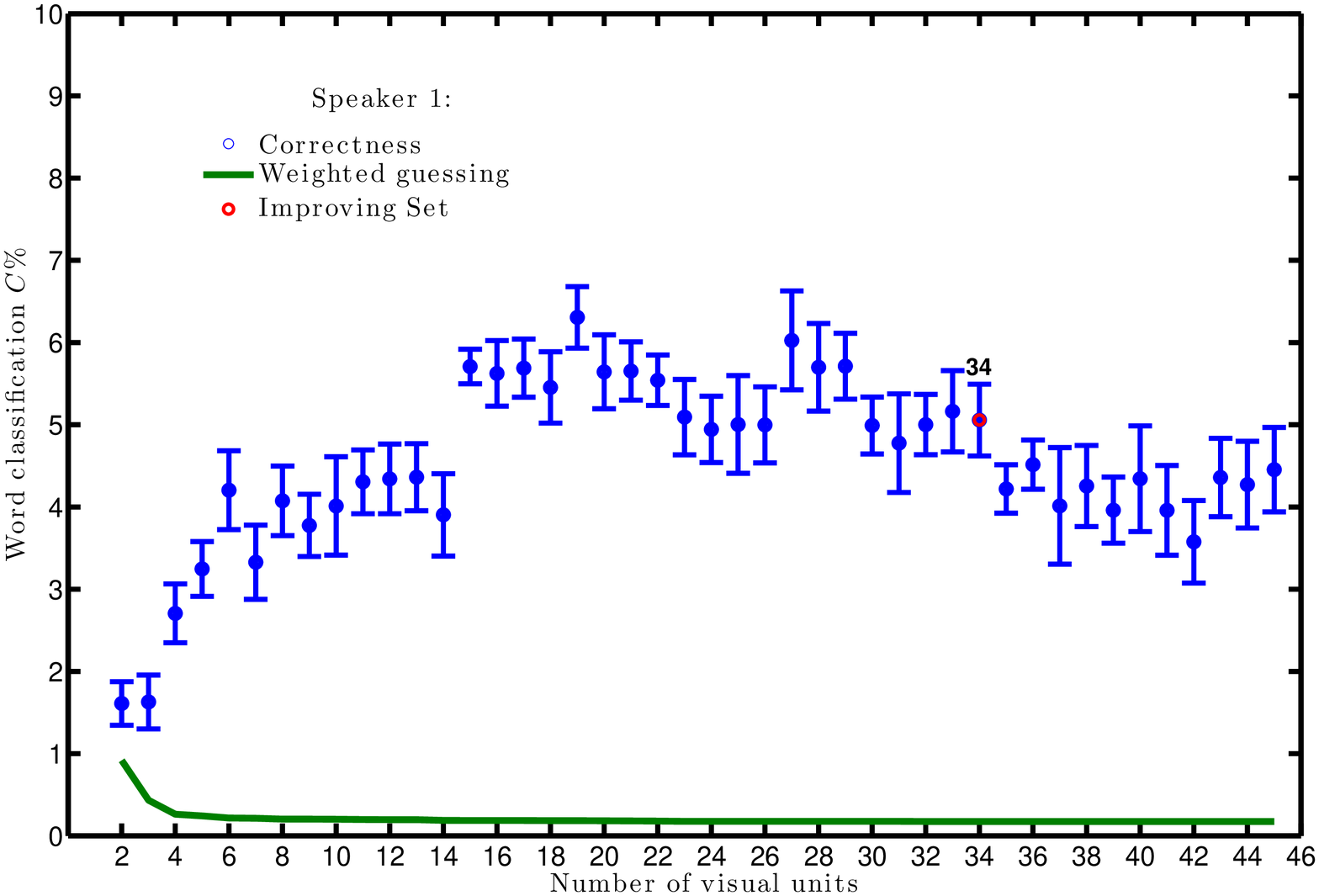} \includegraphics[width=0.49\textwidth]{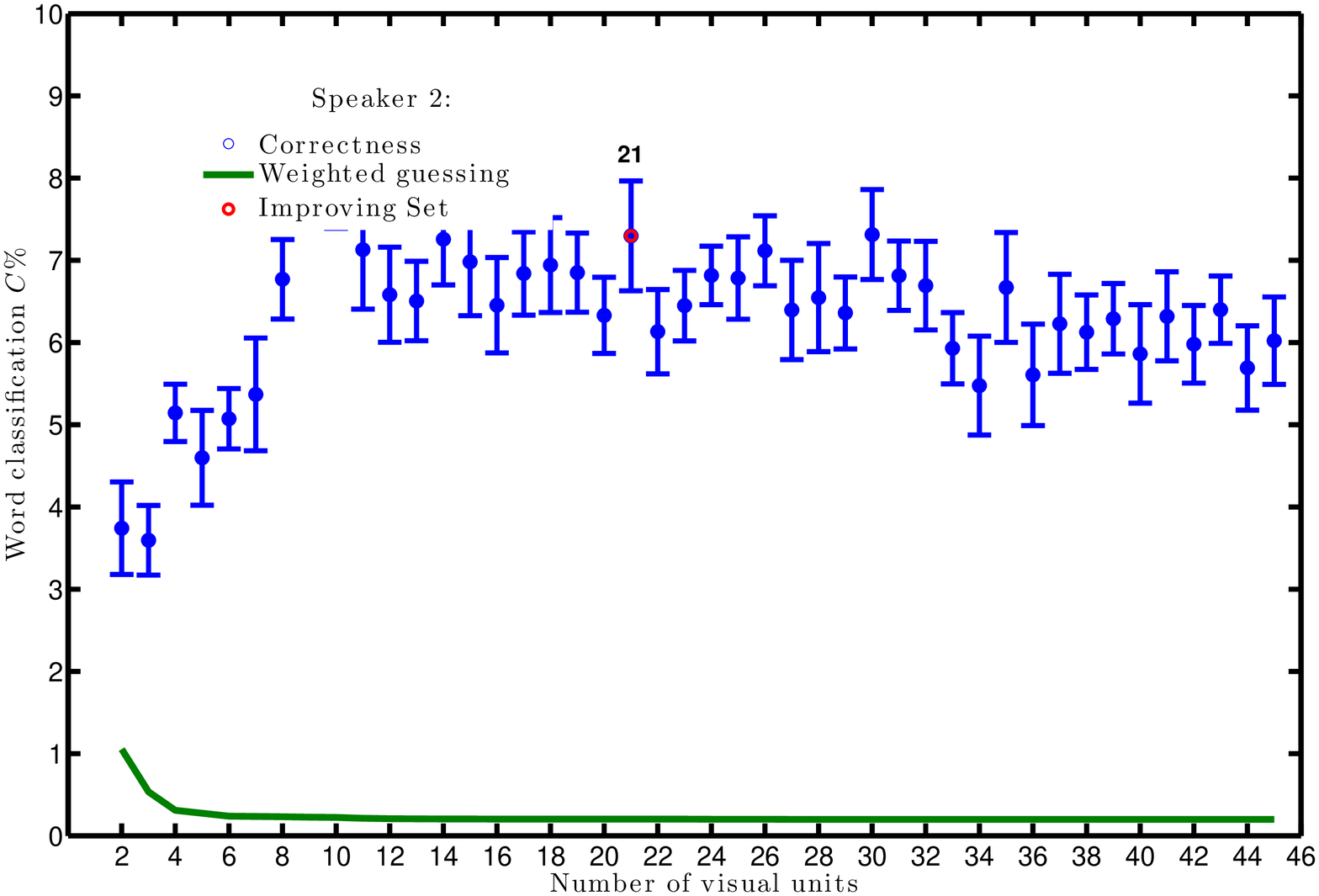} 
\caption{Speaker 1 (\textbf{left}) and Speaker 2 (\textbf{right}): word correctness, $C\pm1se$ for P2V map sizes 2--45. Set 34 (red) is significantly better than set~35.} 
\label{fig:sp01} 
\end{figure}
\unskip 

\begin{figure}[H] 
\centering 
\includegraphics[width=0.49\textwidth]{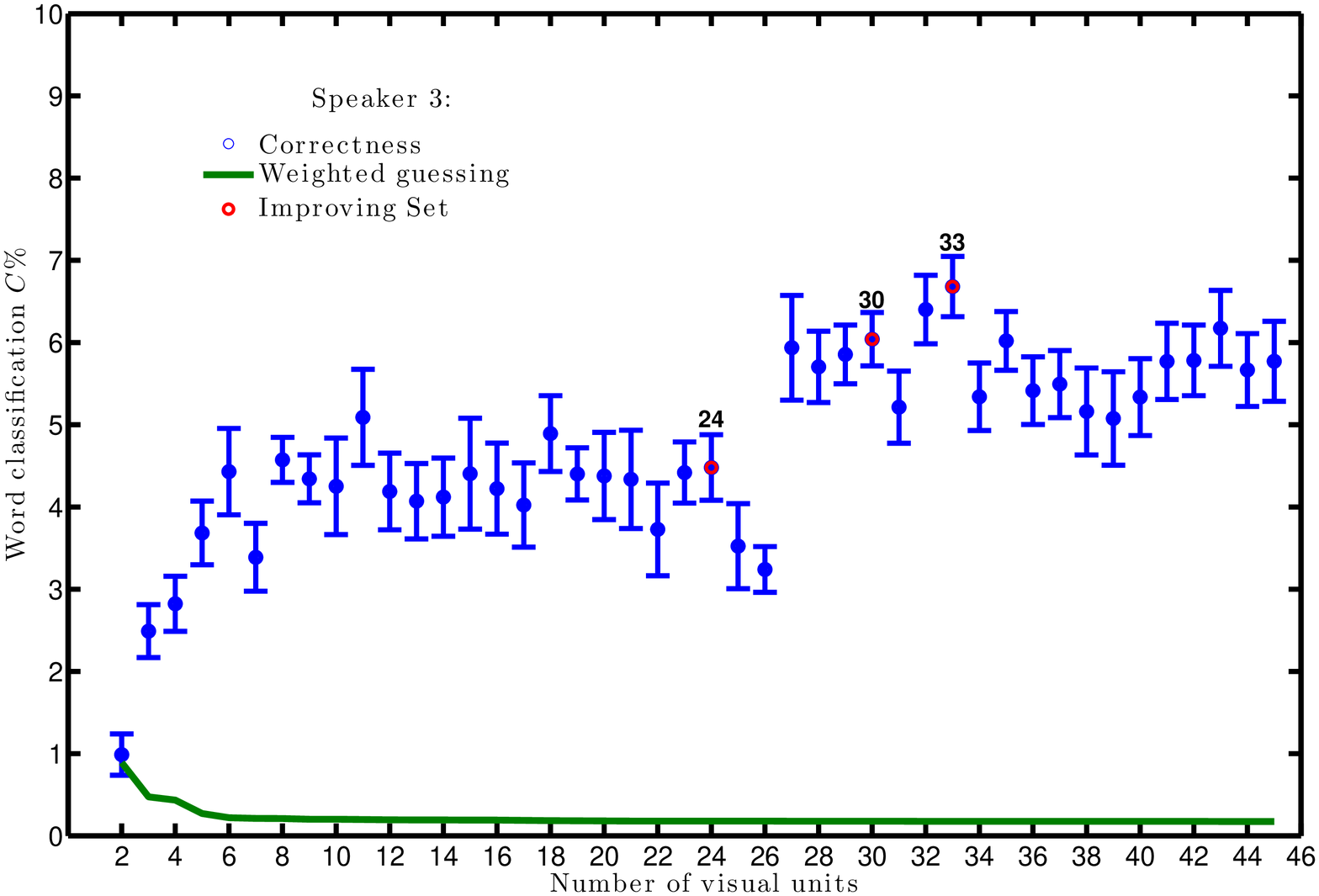} \includegraphics[width=0.49\textwidth]{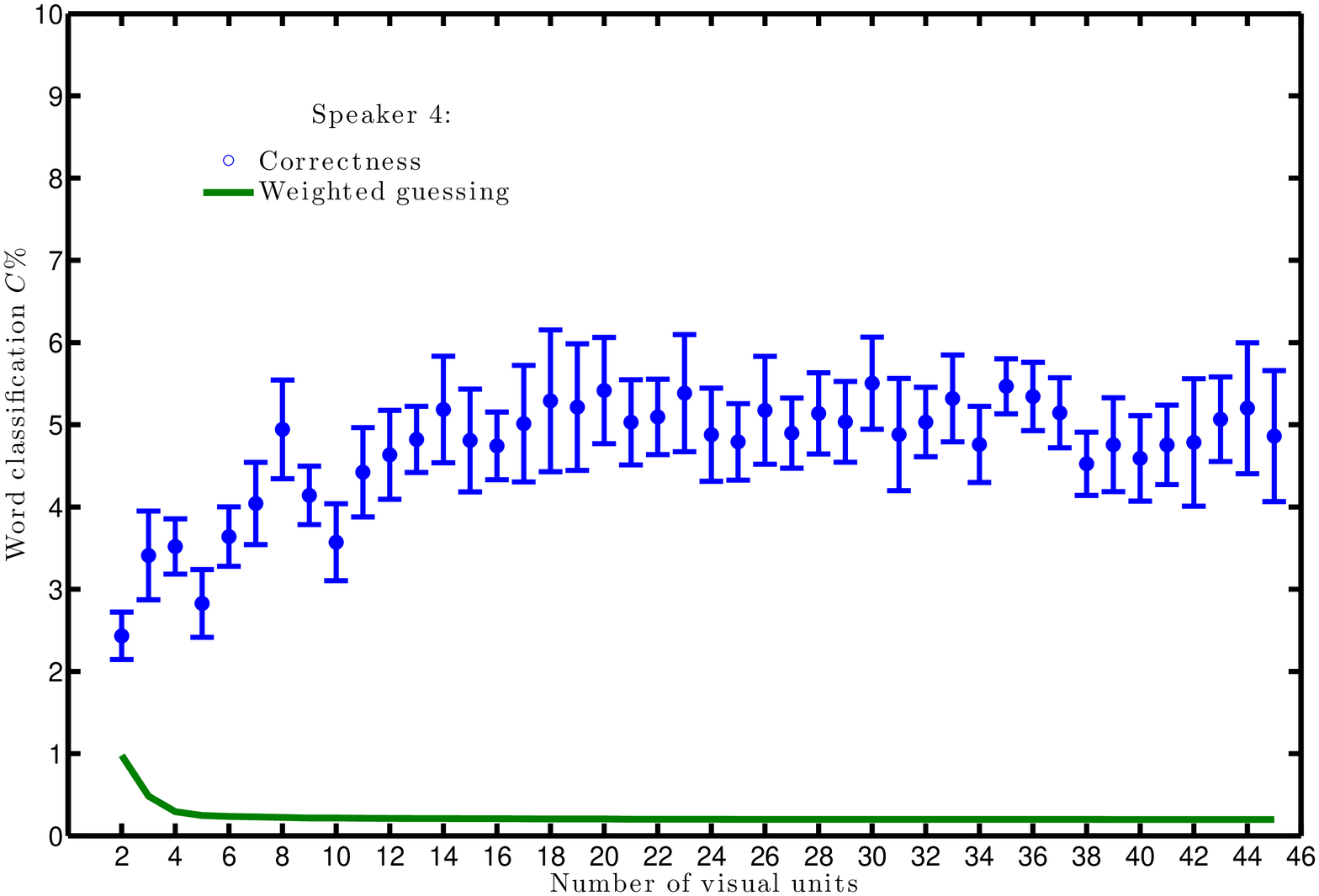} 
\caption{Speaker 3 (\textbf{left}) and Speaker 4 (\textbf{right}): word correctness, $C\pm1se$ for P2V map sizes 2--45. Sets 24, 30, and~33  (red) are significantly better than sets 25, 31, and~34~respectively.} 
\label{fig:sp03} 
\end{figure}
\unskip 

\begin{figure}[H] 
\centering 
\includegraphics[width=0.49\textwidth]{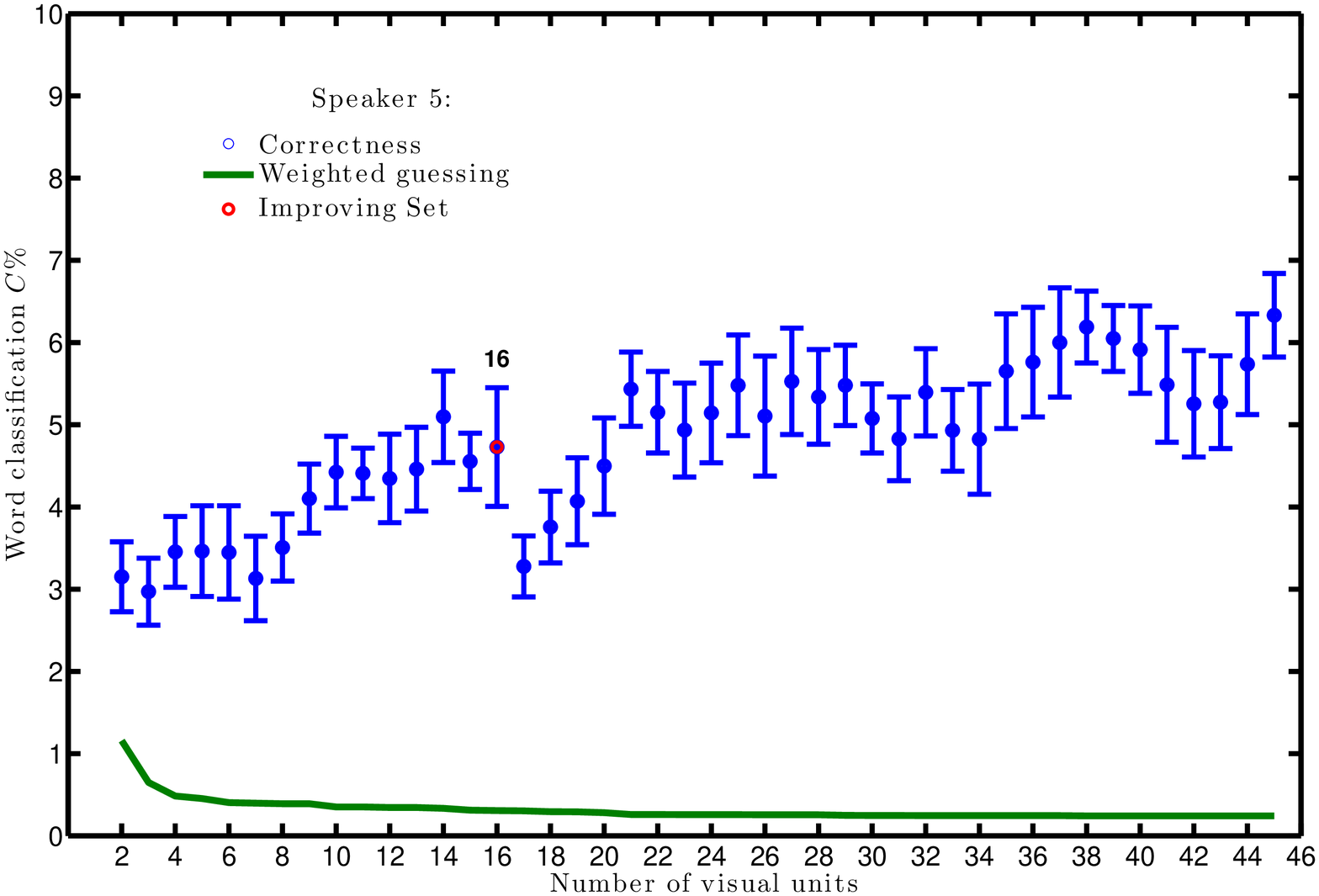} \includegraphics[width=0.49\textwidth]{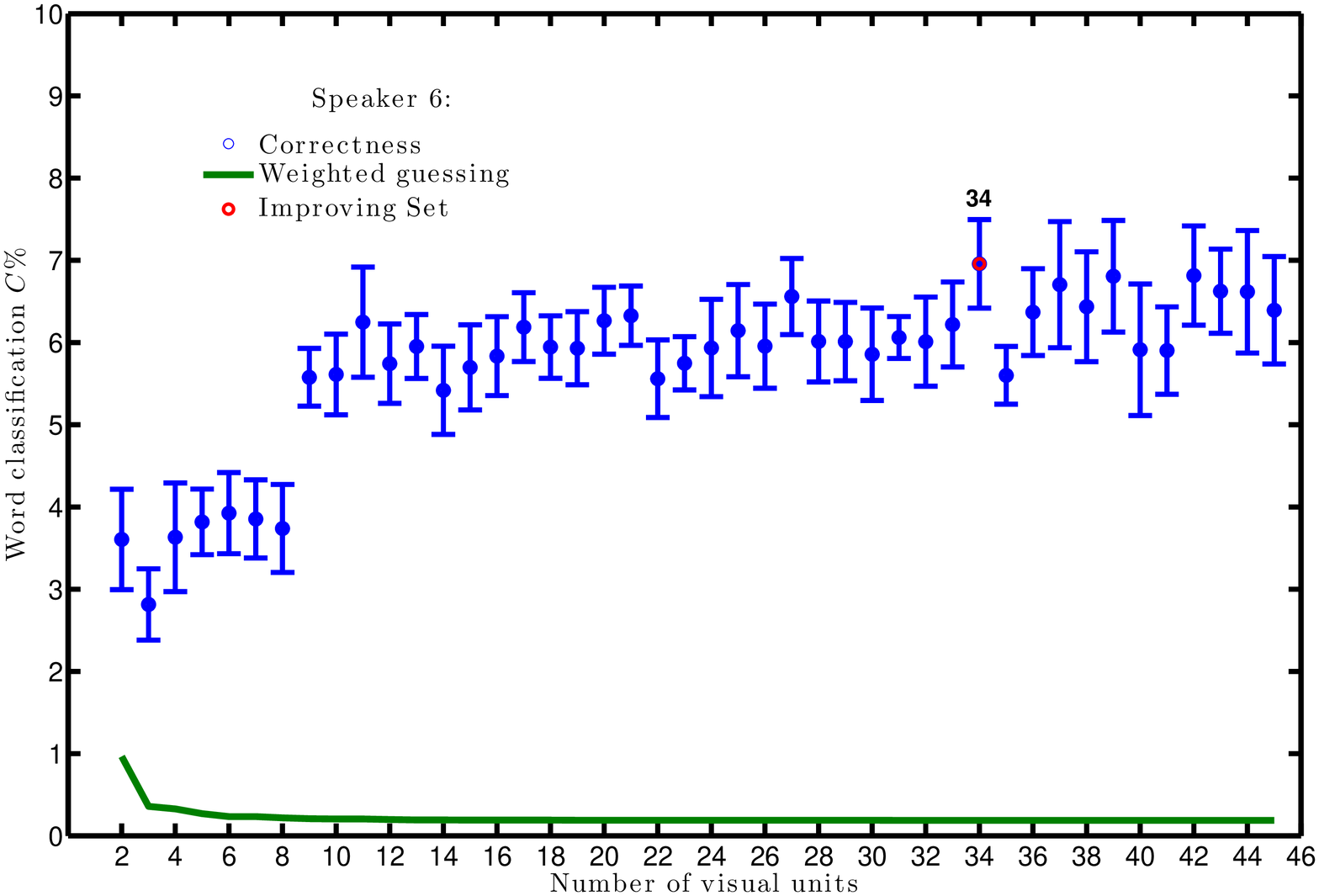} 
\caption{Speaker 5 (\textbf{left}) and Speaker 6 (\textbf{right}): word correctness, $C\pm1se$ for P2V map sizes 2--45. Set 16 (red) is significantly better than set~17.} 
\label{fig:sp05} 
\end{figure}
\unskip 

\begin{figure}[H] 
\centering 
\includegraphics[width=0.49\textwidth]{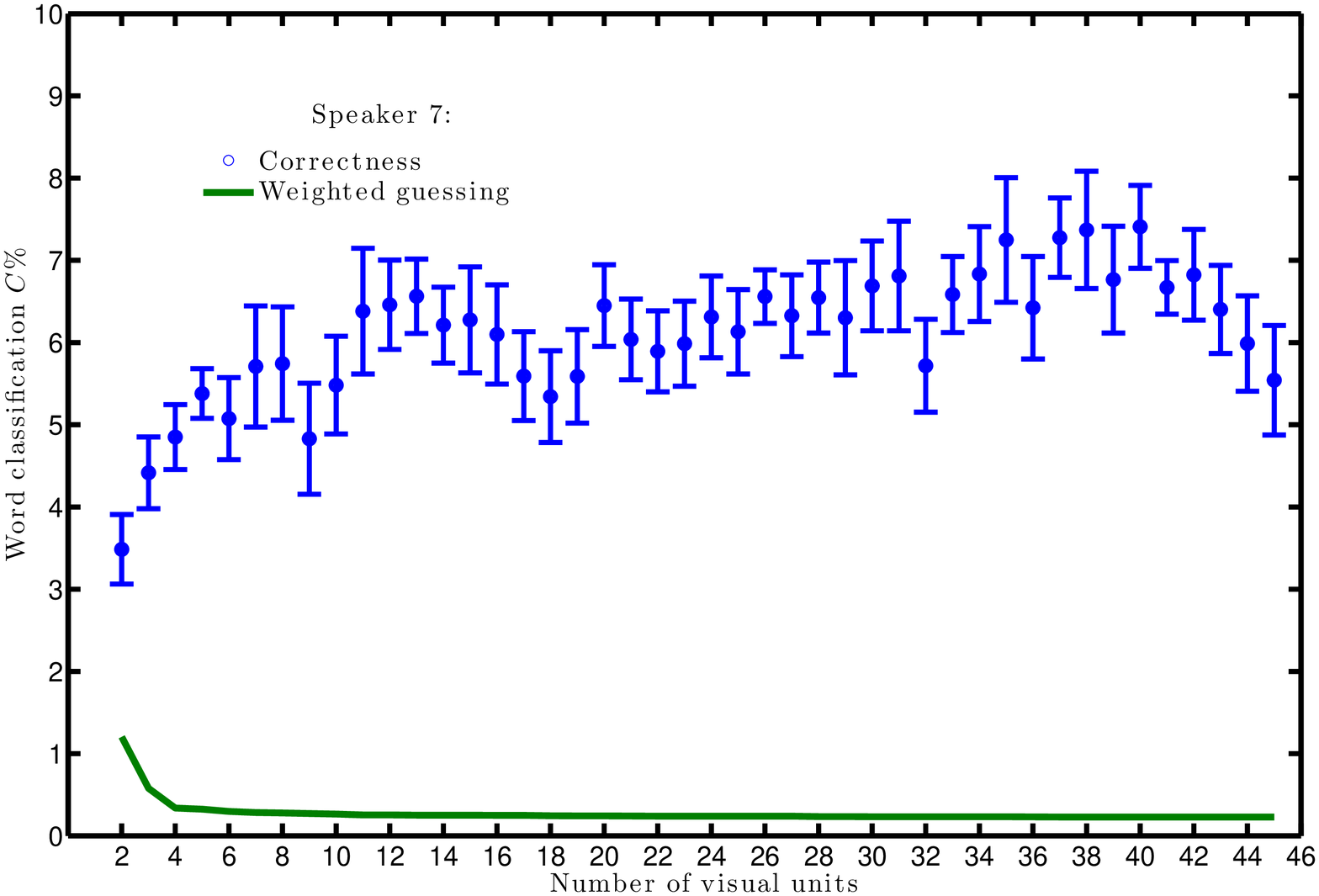} \includegraphics[width=0.49\textwidth]{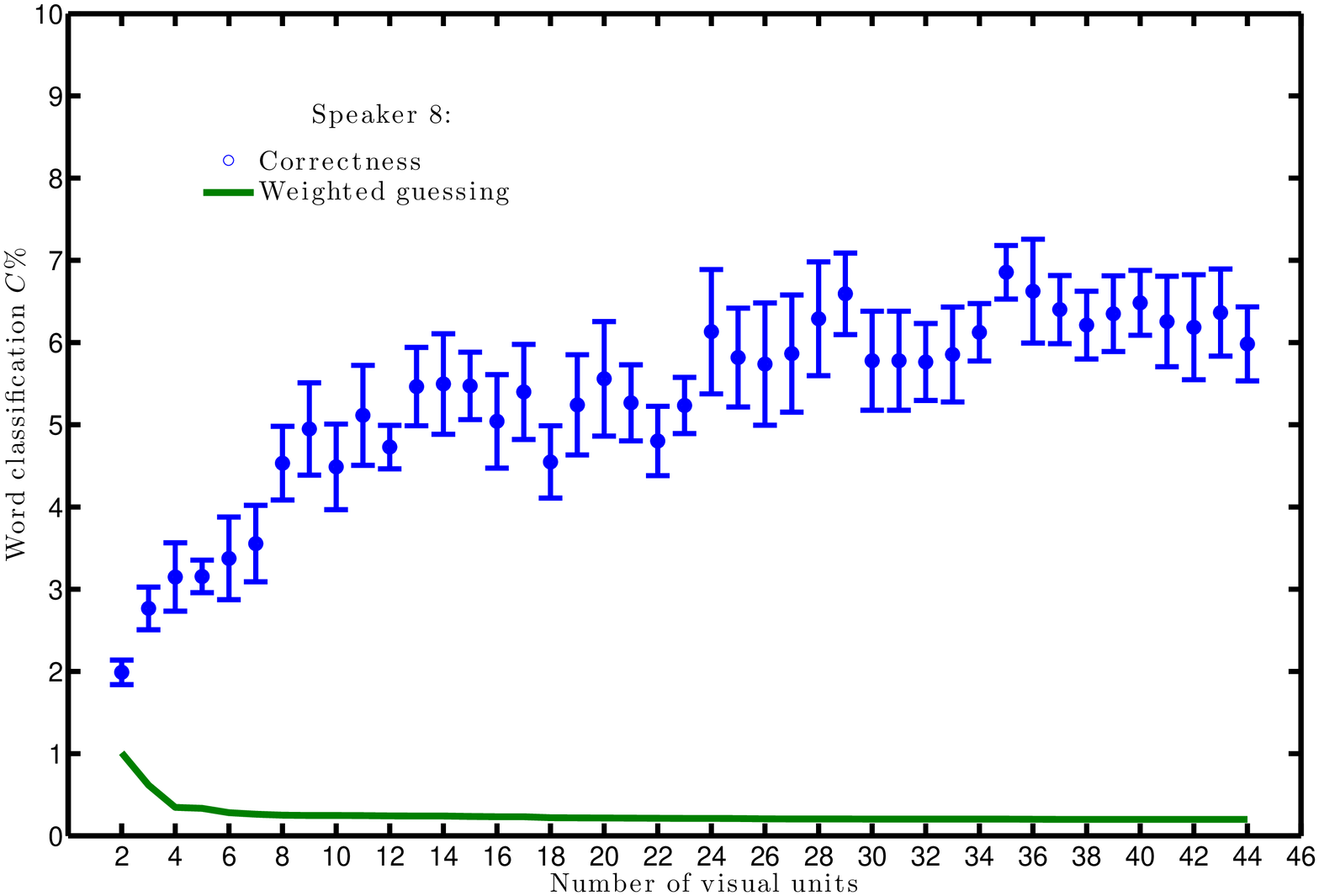} 
\caption{Speaker 7 (\textbf{left}) and Speaker 8 (\textbf{right}): word correctness, $C\pm1se$ for P2V map sizes~2--45.} 
\label{fig:sp07} 
\end{figure}
\unskip 

\begin{figure}[H] 
\centering 
\includegraphics[width=0.49\textwidth]{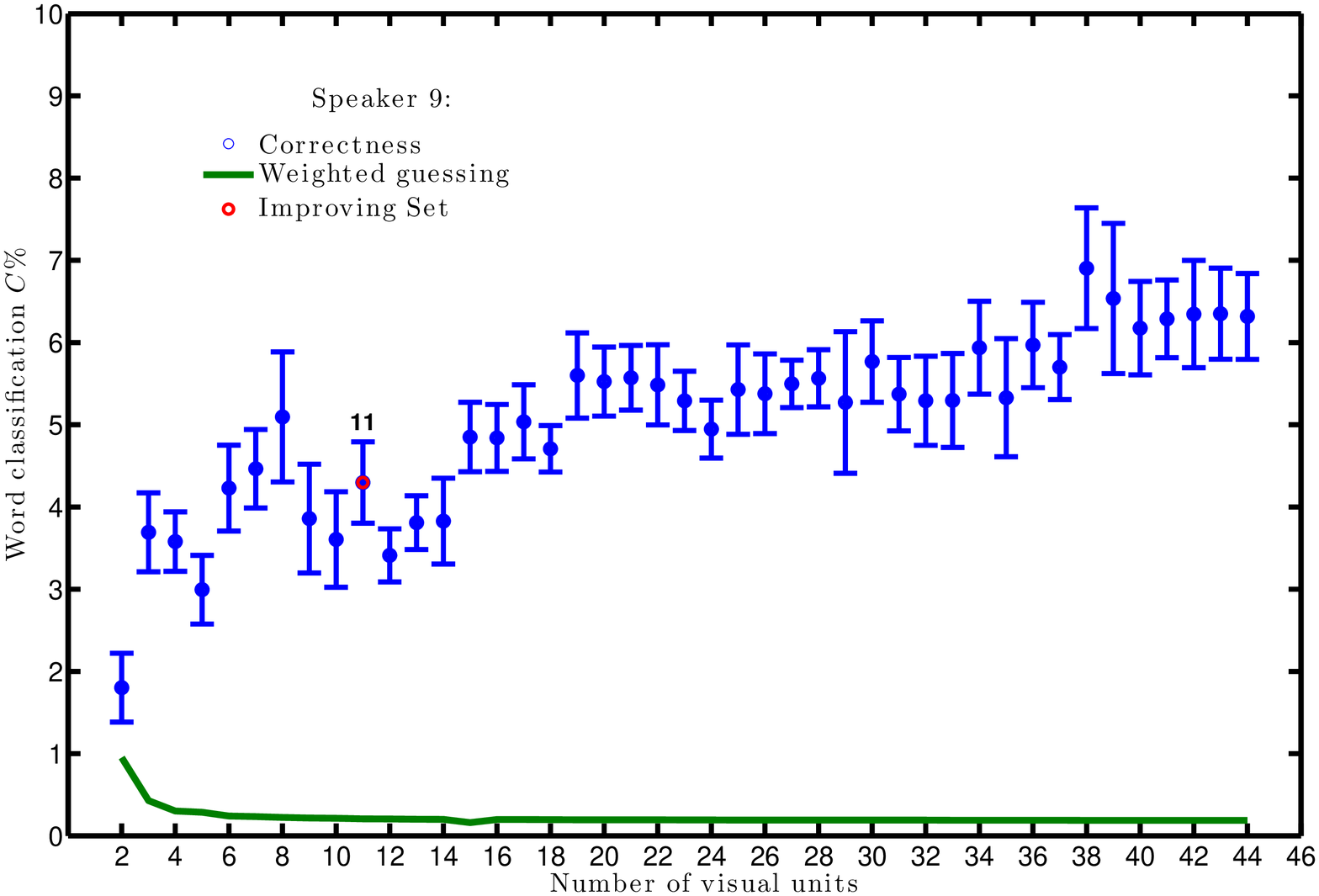} \includegraphics[width=0.49\textwidth]{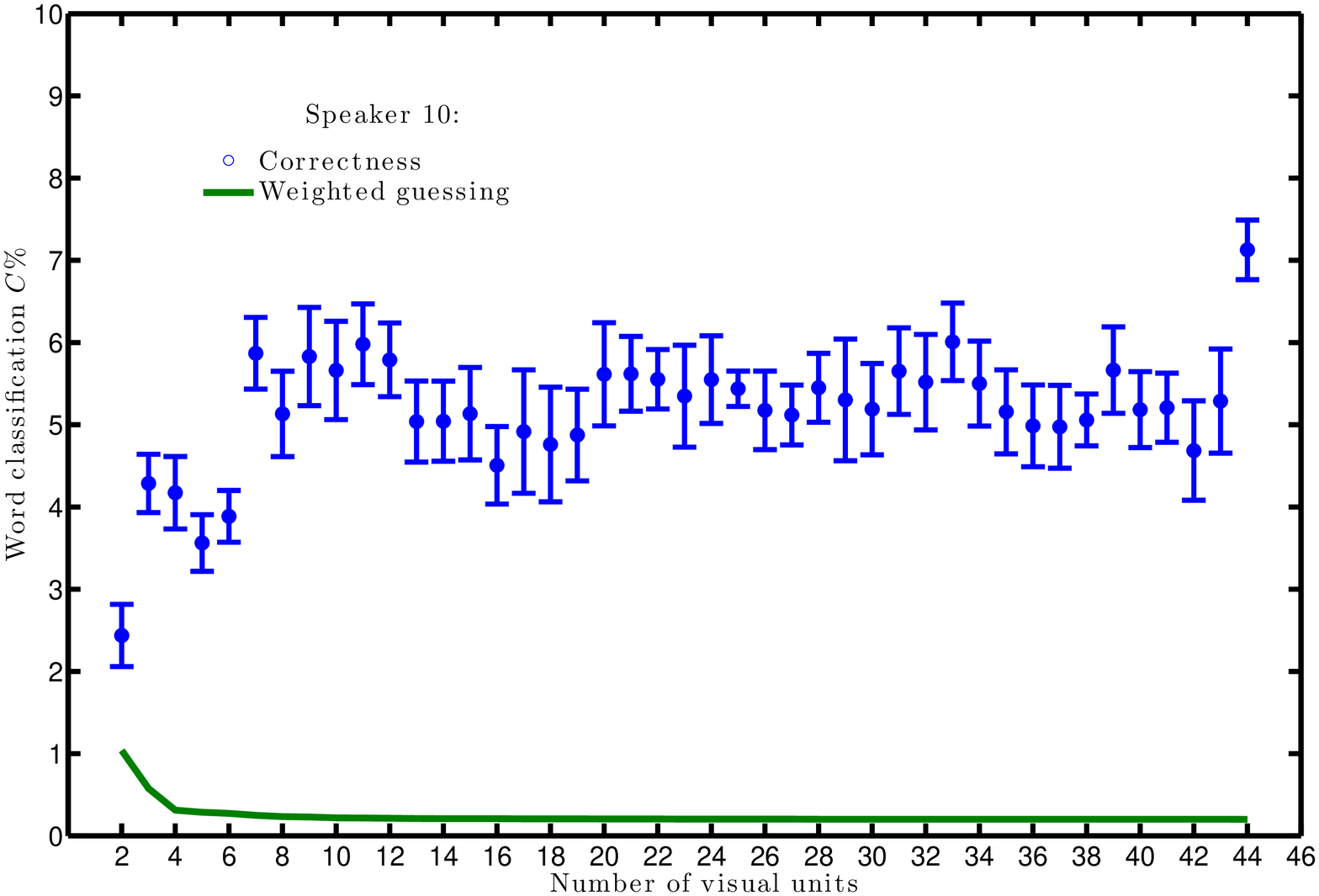} 
\caption{Speaker 9 (\textbf{left}) and Speaker 10 (\textbf{right}) : word correctness, $C\pm1se$ for P2V map sizes 2--44. Set 11 (red) is significantly better than set~12.} 
\label{fig:sp09} 
\end{figure}
\unskip 

\begin{figure}[H] 
\centering 
\includegraphics[width=0.49\textwidth]{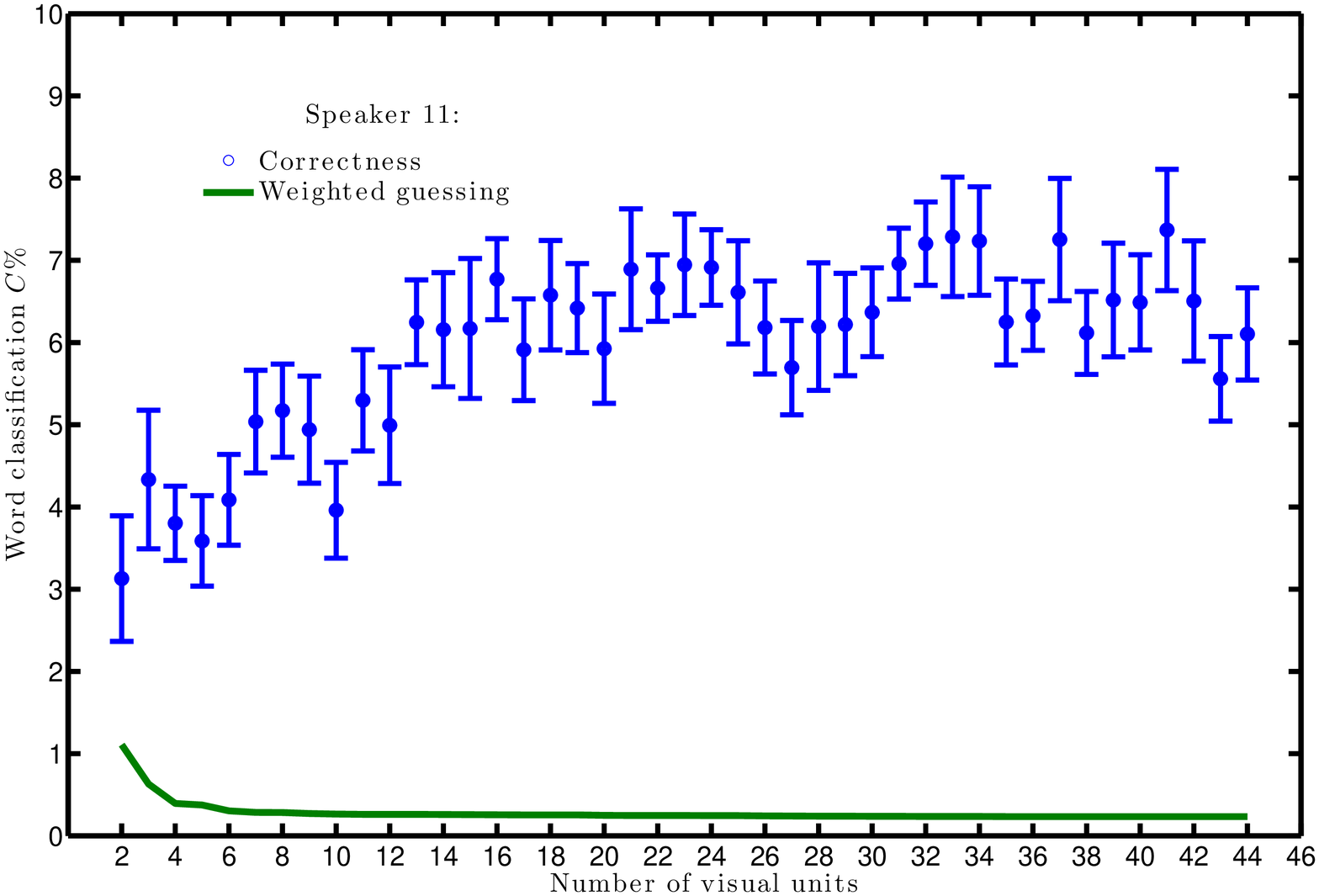} \includegraphics[width=0.49\textwidth]{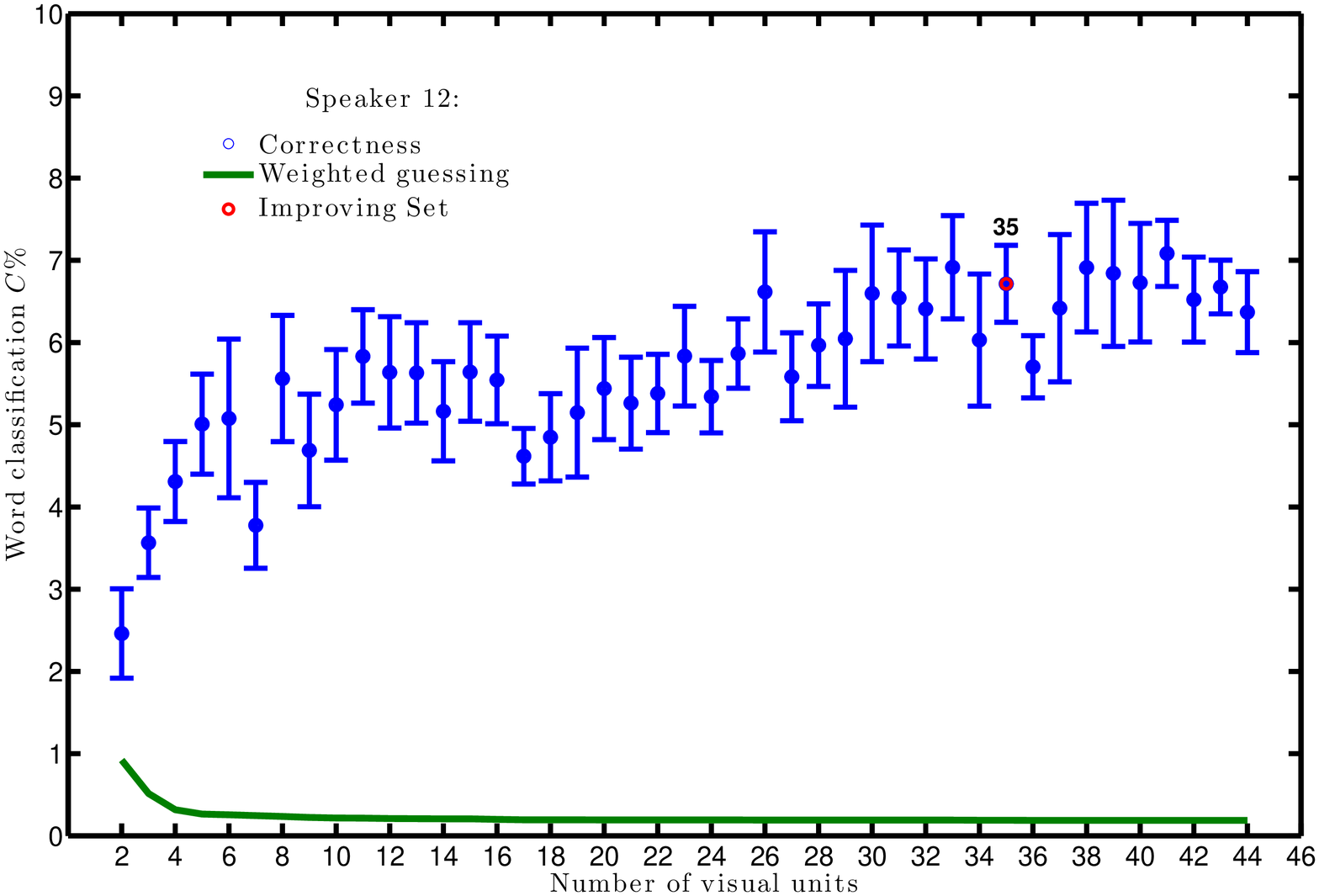} 
\caption{Speaker 11 (\textbf{left}) and Speaker 12 (\textbf{right}): word correctness, $C\pm1se$ for P2V map sizes~2--44.} 
\label{fig:sp11} 
\end{figure} 


\reftitle{References}

\end{document}